\documentclass[11pt]{article}

\usepackage{geometry}
\usepackage{graphicx}
\usepackage{amssymb}
\usepackage{amsmath}
\usepackage{epstopdf}
\newcommand{\nn}{\nonumber}
\newcommand{\be}{\begin{equation}}
\newcommand{\ee}{\end{equation}}
\newcommand{\tr}{\mathrm{tr}}
\newcommand{\diag}{\mathrm{diag}}
\newcommand{\GeV}{~\mathrm{GeV}}
\newcommand{\TeV}{~\mathrm{TeV}}
\newcommand{\MeV}{~\mathrm{MeV}}
\newcommand{\Sec}[1]{Sec.~\ref{#1}}
\newcommand{\Secs}[2]{Secs.~\ref{#1} and \ref{#2}}
\newcommand{\App}[1]{App.~\ref{#1}}
\newcommand{\Fig}[1]{Fig.~\ref{#1}}
\newcommand{\Ref}[1]{Ref.~\cite{#1}}
\newcommand{\Eq}[1]{Eq.~(\ref{#1})}
\newcommand{\Eqs}[2]{Eqs.~(\ref{#1}) and (\ref{#2})}

\newcommand{\ie}{\textit{i.e.}\ }

\title{The Dark Top}

\author{David Poland$^{1,2,3}$, Jesse Thaler$^{1,2}$ \\ \\
\textit{$^{1}$ \small Berkeley Center for Theoretical Physics,
  University of California, Berkeley, CA 94720} \\
\textit{$^{2}$ \small Theoretical Physics Group, Lawrence Berkeley
  National Laboratory, Berkeley, CA 94720}\\
\textit{$^{3}$ \small Jefferson Laboratory, Physics Department, Harvard University, Cambridge, MA 02138}\\
}

\date{}

\begin{document}

\begin{titlepage}

\maketitle
\thispagestyle{empty}

\begin{abstract}
We present a class of composite Higgs models in which the particle that regulates the top quark contribution to the Higgs potential is also a weakly-interacting dark matter candidate.  This color-neutral ``dark top'' is related to the standard model top quark through a large global symmetry.   Because the same couplings that control the Higgs potential also determine various dark matter cross sections, the dark top scenario is quite predictive once the dark top mass and various quantum numbers are specified.  We construct two concrete examples of dark top models with plausible UV completions and study their dark matter properties and LHC signatures.
\end{abstract}

\end{titlepage}

\tableofcontents

\vspace{.3in}

\section{Motivation}

In the standard model (SM), the radiative corrections to the Higgs boson mass parameter are quadratically divergent.  To regulate these divergences in a natural way,\footnote{For a recent discussion see \cite{naturalness}.} the scale at which new physics enters should not be far from the TeV scale.  However, precision tests at LEP \cite{LEP:2005em} suggest that dimension six operators that contribute to electroweak parameters should generically be suppressed by the multi-TeV scale.  Pushing up the scale of new physics to accommodate the LEP constraints while simultaneously keeping the electroweak scale fixed introduces fine-tuning in the Higgs potential, and nearly all theories of physics beyond the standard model (BSM) exhibit some tension between experimental constraints and naturalness.  This tension is present in both supersymmetric and non-supersymmetric theories and is often referred to as the ``LEP Paradox'' \cite{Barbieri:2000gf} or the ``Little Hierarchy Problem'' \cite{Cheng:2003ju}.

There are two well-known techniques to ease the little hierarchy tension.   Recall that weakly coupled BSM theories generically require a partner particle for each SM field, and these partner particles regulate the Higgs quadratic divergences \cite{Dimopoulos:1981zb,Kaplan:1983fs,Kaplan:1983sm,Arkani-Hamed:2000hv,Arkani-Hamed:2001nc,Contino:2003ve,Chacko:2005pe}.   Because the largest radiative correction comes from the top loop, one way to reduce the tension is to make the top partner somewhat lighter than the other partner particles \cite{Cohen:1996vb}.  A second way to ease the tension is to introduce a $\mathbf{Z}_2$ symmetry such as $T$-parity \cite{Cheng:2003ju, Low:2004xc} under which all new particles are odd.  The new particles then always couple to the SM in pairs, ensuring that contributions to electroweak parameters are loop suppressed.

However, it is not clear whether both these techniques could be used simultaneously.  Exact $T$-parity is na\"{\i}vely incompatible with the top partner being the lightest new state---the lightest $T$-odd particle would be not only stable but also colored, and there are strong constraints on stable colored particles.\footnote{For a review see e.g. \cite{Perl:2001xi}.}  The usual way to arrange for both a light top partner and exact $T$-parity is to ensure that there is at least one $T$-odd color neutral particle lighter than the top partner.  In composite Higgs models, this lightest $T$-odd particle is either a gauge boson partner or a pseudo-Goldstone boson (PGB), and is naturally a weakly-interacting dark matter candidate \cite{BirkedalHansen:2003mpa,Cheng:2003ju, Hubisz:2004ft, Martin:2006ss, Birkedal:2006fz}.

In this paper, we pursue a alternative possibility which is more economical in terms of the low energy spectrum. Through a sufficiently exotic UV symmetry structure, we arrange for the top partner to be color neutral, similar to what happens in certain twin Higgs models \cite{Chacko:2005pe,Barbieri:2005ri,Chacko:2005vw}.\footnote{In these twin Higgs models, there is an additional hidden $SU(3)_C'$ gauge group, so while the top partner is color neutral under the standard model, it still has strong annihilation cross sections because of the hidden color group.  In the dark top scenario, the equivalent of the $SU(3)_C'$ gauge symmetry is broken above the dark top mass so the dark top only has weak annihilation cross sections.}  The top partner, being stable but non-colored, is then itself a dark matter candidate.  We call such a particle a ``dark top''.

The dark top scenario is not only economical but also quite predictive.  In a decoupling limit where the only new particle is the dark top, the same interactions that allow the dark top to regulate the Higgs potential will also play a role in determining the annihilation and direct direction cross sections for this dark matter candidate.  After making a set of discrete choices regarding global symmetries and gauge quantum numbers, the main free parameter is just the dark top mass.  In realistic models, there are often non-decoupling effects that skew the dark matter predictions, and we will consider these effects to determine the viable parameter space.  We will find that the dark top is typically a nearly pure $SU(2)_L$ singlet or doublet and has strong couplings to the Higgs sector, and there are regions of parameter space consistent with standard thermal freezeout as well as direct detection experiments.

The LHC signatures of the dark top are more model dependent, but one can make a few generic statements.  We will see that the dark top is almost always quite heavy ($> m_{\rm top}$), so electroweak production of the dark top is not a viable discovery channel. On the other hand, in order for the dark top to be a top partner, it must be in a multiplet with the top, implying some kind of symmetry that relates colored and non-colored particles.   In UV-complete dark top models, there are often PGBs or heavy spin-1 resonances that transform under this new symmetry, yielding color triplets that can decay to a top and a dark top.   If these new states are kinematically accessible at the LHC, strong pair production of the triplets will yield two tops and two dark tops in the final state, mimicking the classic supersymmetry signal of stop pairs decaying to two tops and two neutralinos.

The paper is organized as follows. In \Sec{sec:darktop}, we discuss the general low energy couplings of a dark top, as well as the conditions that UV completions must satisfy in order to ensure that low energy structure.  We show two concrete models that lead to a dark top in \Sec{sec:models} and discuss possible UV completions in \Sec{sec:uvcomplete}.    We analyze the dark matter properties of the dark top in \Sec{sec:darkmatter} and sketch the LHC signatures in \Sec{sec:lhc}.  We conclude in \Sec{sec:outlook}.

\section{A Generic Dark Top}
\label{sec:darktop}

In order for the dark top to regulate the top contribution to the Higgs potential, it must be a top partner, meaning it is in some kind of multiplet with the top.  Taking the dark top to be a fermion,\footnote{It is in principle possible to build a scalar dark top model, but to put a scalar dark top in a multiplet with the top would require not only a colored/non-colored symmetry but also supersymmetry.  For simplicity of discussion, we will postpone the scalar case to future work, and mention only that the structure of the ``dark stop'' is closely related to that of folded supersymmetry \cite{Burdman:2006tz}.} the only known mechanism to have a fermion partner regulate a fermion loop is to have the Higgs be a pseudo-Goldstone boson \cite{Kaplan:1983fs,Kaplan:1983sm,Arkani-Hamed:2001nc,Contino:2003ve}.\footnote{An alternative language is that of gauge-Higgs unification \cite{Fairlie:1979at,Fairlie:1979zy,Manton:1979kb,Hosotani:1983xw,Hosotani:1988bm}, where the shift symmetry that protects the Higgs mass is related to a higher dimensional gauge symmetry.  From a 4D perspective, both descriptions are equivalent.}  Starting with a global symmetry $G$ that is broken to a subgroup $H$ near the TeV scale, the Higgs boson is contained in a $G/H$ nonlinear sigma field $\Sigma$ \cite{Coleman:1969sm,Callan:1969sn},
\be
\Sigma = e^{i \Pi /f}, \qquad \Pi  = \pi^a X^a,
\ee
where $f$ is the decay constant and $X^a$ are the generators in $G/H$.  At low energies, the fermions effectively fill out incomplete multiplets of $G$, so coupling the fermions to $\Sigma$ will generate Yukawa couplings to the Higgs as well as a radiative potential for $\Sigma$.

Generically, the potential $V(\Sigma)$ will be quadratically sensitive to the cutoff, but in certain classes of composite Higgs models, there can be cancellations between the top contribution and the top partner contribution that lead to only logarithmic sensitivity at one loop.  In little Higgs models, these cancellations are the result of collective breaking \cite{Arkani-Hamed:2001nc,Arkani-Hamed:2002qx,Arkani-Hamed:2002qy,Low:2002ws,Chang:2003un,Kaplan:2003uc,Schmaltz:2004de}.  In twin Higgs models, these cancellations are the result of discrete symmetries \cite{Chacko:2005pe,Chacko:2005un}.  In generic dark top models, the symmetries that protects the Higgs are less straightforward, so not surprisingly, the robustness of the cancellation is highly dependent on the UV completion of the theory.  We will postpone a discussion of concrete UV completions to \Sec{sec:uvcomplete}, and here only discuss the requirements for the cancellation to occur.

\subsection{Dark Top Yukawa Sector}

At low energies, the Lagrangian contains a Yukawa interaction with the rough form
\be
\label{eq:topyukawa}
\mathcal{L}_{\rm Yukawa} = \lambda_{\rm top} f Q \Sigma Q^c,
\ee
where $\lambda_{\rm top}$ is the top Yukawa coupling, and $Q$ and $Q^c$ contain the left- and right-handed top quark $q$ and $t^c$, respectively.   If both $Q$ and $Q^c$ were in complete representations of the symmetry group $G$, the Higgs would be an exact Goldstone boson as far as this coupling is concerned, and there would be no top Yukawa interaction.  For general incomplete representations, this coupling introduces quadratically divergent contributions to the Higgs mass.  We will be interested in particular incomplete representations $Q$ and $Q^c$ for which the one-loop quadratic divergence is cancelled.\footnote{In the explicit models presented in \Sec{sec:models}, $Q$ will be an incomplete representation but $Q^c$ will be a complete representation.  The reason for this choice is explained in \Sec{sec:tension}.}

More concretely, we can expand out \Eq{eq:topyukawa} as
\be
\label{eq:yukawa}
\mathcal{L}_{\rm Yukawa} = \lambda_{\rm top} \left( q h t^c + f \kappa^{ij} T_i T_j^c + \cdots \right),
\ee
where $T_i$ and $T^c_j$ are $i,j = 1,\ldots N_D$ vector-like partners of the top.  The matrix $\kappa^{ij}$ is model dependent and a function of the Higgs field $h$.  An illustrative example is
\be
\label{eq:kappaexample}
\kappa^{ij} = \delta^{ij} \left(1 - \frac{h^\dagger h}{2f^2} + \cdots \right).
\ee
With this special form of the $\kappa^{ij}$ matrix and the choice $N_D = 3$, we will see that there is a cancellation in the one-loop quadratic divergence.  This special form of the $\kappa^{ij}$ matrix can be guaranteed by the symmetries of the non-linear sigma model at tree-level and we will present examples of such symmetries in \Sec{sec:models}.  In fact, the $\kappa^{ij}$ matrix in \Eq{eq:kappaexample} is familiar from certain little Higgs models \cite{Kaplan:2003uc, Cheng:2005as}, and in those contexts $T$ is a color triplet.  However, there is no need for these partners to be colored as long as $N_D = 3$.

The fermions $T$, $T^c$ are dark matter candidates because they can be made odd under an exact $\mathbf{Z}_2$ symmetry.   The spectrum of these dark tops depends on the specific form of $\kappa^{ij}$, but in all cases, the masses of the dark tops will be tied to the mass of the top, leading to
\be
\frac{m_i}{m_{\rm top}} \simeq c_i \frac{f}{v} + \mathcal{O}(v^2/f^2),
\ee
where the Higgs vev $v = \langle h \rangle \simeq 175 \GeV$ and $c_i$ is an $\mathcal{O}(1)$ number that is in general different for the three $T_i$ eigenstates.  By assumption $f > v$ in composite Higgs models, so we see that the dark tops are typically heavier than the top quark, expect when $f$ is very close to $v$.

\subsection{One-Loop Cancellation}

\begin{figure}
\begin{center}
\includegraphics[scale=0.35]{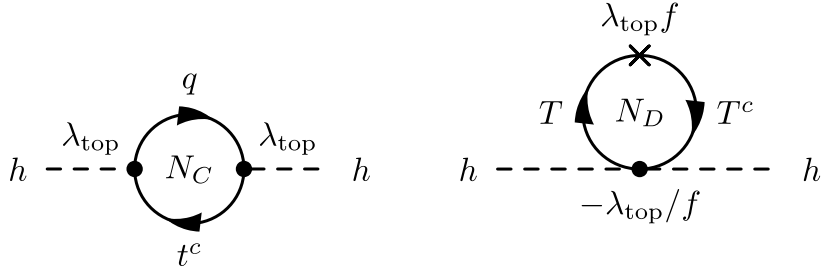}
\end{center}
\caption{Schematic cancellation of the quadratic contribution to the Higgs mass at one-loop between the $N_C = 3$ species of tops and the $N_D = 3$ species of dark tops.  Because of a large global symmetry, the dark top interactions and the top Yukawa coupling are determined by the same parameter $\lambda_{\rm top}$.  At one-loop it is not necessary for the tops and dark tops to have the same $SU(3)_C$ quantum numbers, allowing for the dark tops to be weakly-interacting dark matter candidates.  Whether an effective cancellation persists beyond the one-loop level depends on the details of the UV completion.
\label{fig:cancellation}}
\end{figure}

For special forms of $\kappa^{ij}$, the quadratically divergent one-loop contribution to the Higgs potential from the $N_C=3$ species of top quarks will cancel against loops of the $N_D$ species of dark tops as in \Fig{fig:cancellation}.   To see more explicitly why there is a cancellation, consider the Coleman-Weinberg potential \cite{Coleman:1973jx} in the fermion sector about a fixed Higgs background
\be
\label{eq:colemanweinberg}
V(v) = -\frac{1}{8\pi^2} \tr \left[M^\dagger M \Lambda^2 \right] + \frac{1}{8\pi^2} \tr \left[ (M^\dagger M)^2 \log \frac{\Lambda^2}{M^\dagger M}  \right],
\ee
where $M \equiv M(v)$ is the Dirac mass matrix of all fermions and $\Lambda$ is the UV cutoff.  The mass matrix for the $N_C = 3$ colors of top quarks is
\be
M_{\rm top} = \lambda_{\rm top} (v + \cdots) \left(\begin{array}{ccc}1 & 0 & 0 \\0 & 1 &0 \\0 & 0 &1 \end{array}\right), \qquad \tr \left[ M^\dagger_{\rm top} M_{\rm top} \right] = \lambda_{\rm top}^2 N_C v^2 + \cdots.
\ee
Taking \Eq{eq:kappaexample} as an example, the mass matrix for the $N_D = 3$ flavors of dark top is
\be
M_{\rm dark} = \lambda_{\rm top} \left(f - \frac{v^2}{2f} + \cdots \right)\left(\begin{array}{ccc}1 & 0 & 0 \\0 & 1 &0 \\0 & 0&1 \end{array}\right), \qquad  \tr \left[ M^\dagger_{\rm dark} M_{\rm dark} \right] = \lambda_{\rm top}^2 N_D \left(f^2 - v^2 + \cdots \right).
\ee
Summing these contributions, the quadratically divergent piece of the Coleman-Weinberg potential is proportional to
\be
\label{eq:trM2cancel}
\qquad \tr \left[ M^\dagger_{\rm total} M_{\rm total} \right] = \mbox{const} + \lambda_{\rm top}^2 (N_C - N_D)v^2 +\cdots,
\ee
so for $N_C = N_D$, the divergent top contribution to the Higgs mass parameter vanishes.

In writing down the Yukawa couplings in \Eq{eq:yukawa}, we did not need to specify the quantum numbers of $T$, $T^c$.  All three $T_i$ can be electroweak singlets, or there can be one $SU(2)_L$ doublet and one singlet.  Depending on the quantum numbers of the dark top fields, different choices for $\kappa^{ij}$ are possible.  An example we will encounter in \Sec{sec:simplegroup} is
\be
\label{eq:kappaexamplealt}
\kappa^{ij} = \left(\begin{array}{ccc}c_v & 0 & s_v \\0 & c_v &0 \\0 & 0 & \sqrt{2} c_v \end{array}\right), \qquad c_v \equiv \cos \frac{v}{f}, \quad s_v \equiv \sin \frac{v}{f},
\ee
which naturally arises if $(T_A, T_B)$ are in an $SU(2)_L$ doublet and $T_C$ is a singlet.  It is straightforward to verify that this choice of $\kappa^{ij}$ does yield the desired Higgs vev dependence:
\be
\tr \left[ M^\dagger_{\rm dark} M_{\rm dark} \right] = \lambda_{\rm top}^2 f^2 \kappa_{ij} \kappa^{ji} = \lambda_{\rm top}^2 f^2 \left(4 c_v^2 + s_v^2  \right)= \mbox{const} - \lambda_{\rm top}^2 N_D^{\rm eff} v^2 + \cdots,
\ee
with $N_D^{\rm eff} = 3$.  There are many different choices for $\kappa^{ij}$ that work, and there can even be more than three dark top fields as long as the dark top Yukawas effectively yield $N_D^{\rm eff} = 3$ in the Coleman-Weinberg potential.

\subsection{UV Considerations}
\label{subsec:uvconsiderations}

From a low energy point of view, any choice of $\kappa^{ij}$ that yields $N_D^{\rm eff} = 3$ is equally valid.  However, there are constraints on UV completions that are not visible from the Coleman-Weinberg potential alone.  In a full theory, the cutoff $\Lambda^2$ should really be thought of as a matrix with different cutoffs for different fermion fields.  The vanishing of $\tr |M|^2$ does not immediately imply the vanishing of $\tr \left[|M|^2 \Lambda^2 \right]$ unless one can find a UV completion for which $\Lambda^2$ is nearly proportional to the identity matrix.  Another way to say this is that if we start with a $G$-symmetric coupling, then only particular ways of decoupling some of the fields will preserve the structure of \Eq{eq:yukawa} at tree-level.  We will present UV completions having this property in \Sec{sec:uvcomplete}.  Also, because the top and dark top have different gauge quantum numbers, one worries about a large quadratic divergence proportional to $g_s^2 \lambda_{\rm top}^2$ appearing at two-loops, but we will argue that the two-loop contribution should be of reasonable size in concrete models.

Unlike in little Higgs theories, we will not discuss the origin of the quartic coupling for the Higgs, and simply assert that the Higgs vev $v$ can be much smaller than the $\Sigma$ decay constant $f$, possibly because of a small fine-tuning in the Goldstone potential.  Also, we are not introducing low energy partners for the standard model gauge bosons, which means that the quadratically sensitive gauge boson contributions to the Higgs potential are unregulated.  In this sense, our framework is similar to that of the intermediate Higgs \cite{Katz:2005au}. In a UV-complete context, we will find that the size of the radiative contribution to the Higgs potential scales like
\be
\label{eq;gaugeovertop}
\frac{|\delta m_h^2|_{\rm gauge}}{|\delta m_h^2|_{\rm top}} \sim \frac{g_{\rm EW}^2 \Lambda_{\rm TC}^2}{N_C \lambda_{\rm top}^2 f^2},
\ee
where $\Lambda_{\rm TC}$ sets the scale of heavy spin-one resonances.  To the extent that this ratio can be made order one with a suitable choice for $\Lambda_{\rm TC}$, it makes sense for the top loop to be regulated at a parametrically lower scale than the gauge loop.

\begin{figure}
\begin{center}
\includegraphics[scale=0.65]{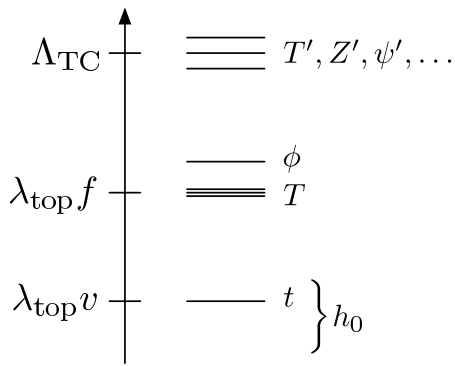}
\end{center}
\caption{The typical spectrum of a dark top model.  For the dark top scenario to be well-defined, $v < f < \Lambda_{\rm TC}$.  The top and Higgs masses are around $\lambda_{\rm top} v$.  The dark tops are at $\lambda_{\rm top} f$, with possible mass splittings of order $\lambda_{\rm top}f$, $\lambda_{\rm top}v$, or $\lambda_{\rm top} v^2 / f$ depending on the symmetry structure.   There can be additional colored PGBs $\phi$ around (but heavier than) $\lambda_{\rm top} f$.  The fields $T'$ around $\Lambda_{\rm TC}$ regulate the one-loop logarithmic divergent and two-loop quadratic divergent contributions to the Higgs potential from tops and dark tops.  Additional gauge boson and fermion partners are also around $\Lambda_{\rm TC}$.  We assume that $\Lambda_{\rm TC} \simeq 4 \pi f$ for our discussion, though $\Lambda_{\rm TC}$ could be smaller depending on the UV completion.
\label{fig:spectrum}}
\end{figure}

The generic spectrum of the dark top scenario is shown in \Fig{fig:spectrum}.  The physical Higgs mass is comparable to the top quark mass $\lambda_{\rm top} v$.  The dark tops are at $\lambda_{\rm top} f$ and may be accompanied by additional PGBs $\phi$ around (but assumed heavier than) that scale.  The remaining standard model partner particles are at the scale $\Lambda_{\rm TC}$, as well as the fields $T'$ which define the cutoff $\Lambda$ in the dark top sector.  In typical composite Higgs models, the mass of $W'/Z'$-like states that mix with electroweak gauge bosons have to be heavier than around 2 TeV in order to satisfy precision electroweak tests \cite{Csaki:2002qg,Agashe:2005dk}.  In the dark top, these gauge bosons appear at the scale $\Lambda_{\rm TC}$, but because the ratio of $f$ to $\Lambda_{\rm TC}$ is model dependent, we cannot put a bound directly on $f$.  Scaling up the ratio $m_\rho/f_\pi \sim 6.5$ from QCD gives a rough bound $f > 300 \GeV$, but we will not impose this constraint on $f$ in this paper.

\section{Two Concrete Models}
\label{sec:models}

Before we can discuss the dark matter properties of the dark top, we have to introduce some concrete models.  The reason is that if the relic abundance is determined by standard thermal freezeout, then the annihilation channel $T \overline{T} \rightarrow W^+ W^-$ or the coannihilation channel $T \overline{T'} \rightarrow W^* \rightarrow \psi \overline{\psi'}$ can play a dominant role in determining the $SU(2)_L$ dark doublet abundance.  Therefore, we need concrete models to determine whether our dark matter candidate is dominantly $SU(2)_L$ doublet or dominantly $SU(2)_L$ singlet.

Since the Higgs comes from the symmetry breaking pattern $G/H$, the standard model gauge groups must be somehow embedded in $G$.  In particular, we will gauge a subgroup $F$ in $G$, and the standard model gauge bosons will be contained in $F \cap H$.   Considering just $SU(3)_C$ and $SU(2)_L$, there are two broad categories of dark top models one can build, though more exotic structures are also possible. Product group models have $G = G_C \times G_L$, with $SU(3)_C \subset G_C$ and $SU(2)_L \subset G_L$.  Simple group models have both $SU(3)_C$ and $SU(2)_L$ as subgroups of a single simple group $G$.

We will focus on one concrete example from each of these classes:  a product group example with $SU(2)_L$ singlet dark matter, and a simple group example with $SU(2)_L$ doublet dark matter.  We will then discuss why a maximal doublet/singlet mixing scenario is disfavored by the dark top requirements.  For simplicity, we will talk about a linear sigma field $\Phi$ instead of a non-linear sigma field $\Sigma$ when discussing the Yukawa couplings.

\subsection{Product Group Example}
\label{sec:productgroup}

One example of a product group dark top comes from the symmetry structure
\begin{align}
F &= SU(3)_C \times SU(2)_L, \nn\\
G &= SU(6)_C \times SU(3)_L, \nn\\
H &= SU(6)_C \times SU(2)_L,
\end{align}
where $F$ is the gauged subgroup of $G$, and the global symmetry breaking pattern is $G/H$.  $SU(3)_C$ is the ordinary color group, and $SU(2)_L$ is the ordinary weak group.  We will temporarily turn off hypercharge\footnote{To get the right hypercharge assignment, one needs to introduce additional $U(1)$ gauge fields, as in Refs.~\cite{Kaplan:2003uc,Contino:2003ve,Thaler:2005en}.  The simplest way to think about hypercharge is to treat $Q$, $Q^c$, and $\Phi$ as transforming under $U(6)_C \times U(3)_L$.   In this notation, the hypercharge generator is $Y = Y_C + Y_L$ where $Y_C = \diag(2/3,2/3,2/3,0,0,0)$ is a generator of $U(6)_C$ and $Y_L = \diag(1/2,1/2,0)$ is a generator of $U(3)_L$.} and the non-top standard model fermions for simplicity.

To break $G \rightarrow H$, consider a linear sigma model field $\Phi$ that transforms as a $(\mathbf{1},\mathbf{3})$ under $G$ and implements the breaking pattern $SU(3)_L/SU(2)_L$.  This yields a pseudo-Goldstone boson doublet and singlet in $\Phi$
\be
\Phi = e^{i \Pi / f}\left(\begin{array}{c}0 \\0 \\f\end{array}\right), \qquad \Pi = \left(\begin{array}{ccc}0 &  0 & h_1 \\0 & 0 & h_2 \\h_1^\dagger & h_2^\dagger & 0 \end{array}\right) + \frac{\eta}{2 \sqrt{2}} \left(\begin{array}{ccc}1 &  0 & 0 \\0 & 1 & 0 \\ 0& 0 & -2 \end{array}\right),
\ee
where $(h_1,h_2) = h$ is the ordinary complex Higgs doublet, and $\eta$ is a real singlet.  Note that in principle, $\eta$ could have been the dark matter candidate because $\eta$ is typically lighter than the $T_i$ fields.  However, in \Sec{sec:uvcomplete} we will gauge the generator corresponding to the $\eta$ direction in order for $\eta$ to be eaten by a heavy gauge boson and lifted above the dark top.

Under the symmetry $G$, there is a $Q$ fermion that transforms as $(\mathbf{6},\mathbf{\overline3})$ and a $Q^c$ that transforms as $(\mathbf{\overline6},\mathbf{1})$.  We can write down the Yukawa coupling
\be
\label{eq:Yukawa63}
\mathcal{L}_{\rm Yukawa} = \lambda_{\rm top} Q \Phi Q^c.
\ee
Decomposing $Q$ and $Q^c$ as
\be
\label{eq:decompose63}
Q = \left(\begin{array}{cccccc}q_{1r} & q_{1g} & q_{1b} & 0 & 0 & 0 \\q_{2r} & q_{2g} & q_{2b}  & 0 & 0 & 0 \\0 & 0 & 0 & T_A & T_B & T_C\end{array}\right), \qquad Q^c = \left(\begin{array}{cccccc}t_r^c & t_g^c & t^c_b & T_A^c & T_B^c & T_C^c\end{array}\right),
\ee
where $r$, $g$, $b$ are the usual color indices, we see that \Eq{eq:Yukawa63} reproduces precisely the Yukawa sector of \Eq{eq:yukawa} with the $\kappa_{ij}$ matrix of \Eq{eq:kappaexample}.  An explanation for the symmetries that protect the Higgs are presented in \App{app:symmetries}.

By zeroing out certain components of $Q$ and $Q^c$, we are explicitly violating the $G$ symmetry, but the UV completions we present in \Sec{sec:explicitads} will allow such breaking without spoiling the form of \Eq{eq:yukawa}.  Note that we could have replaced $SU(6)_C$ with just $SU(3)_{C1} \times SU(3)_{C2}$ with a $\mathbf{Z}_2$ interchange symmetry.  The reason to consider $SU(6)_C$ is the possibility of composite color triplet spin-one fields.  We will return to the field content in \Sec{subsec:uvfield}.

In order to talk about moderate values of $v/f$, it is worth keeping track of $v/f$ corrections to various physical quantities in terms of
$c_v \equiv \cos \frac{v}{f}$ and $s_v \equiv \sin \frac{v}{f}$. Taking $\langle h \rangle = v$, the physical electroweak vev is
\be
v_{\rm physical} = f s_v,
\ee
so the $W$ boson mass is $g_{\rm EW} v_{\rm physical}$ and the top mass is $\lambda_{\rm top} v_{\rm physical}$.  The canonically normalized physical Higgs boson $h_0$ is parametrized as
\be
h \equiv v + \frac{h_0}{\sqrt{2}}.
\ee
In unitary gauge and only considering terms linear in $h_0$, the $\Phi$ field is parametrized as
\be
\label{eq:phiexpansioninh0}
\Phi = \left(\begin{array}{c}v_{\rm physical} + \frac{c_v}{\sqrt{2}} h_0 \\0 \\ f c_v -  \frac{v_{\rm physical}}{f\sqrt{2}}  h_0  \end{array}\right), \qquad c_v \equiv \sqrt{1 - \frac{v_{\rm physical}^2}{f^2}}.
\ee

As far as the dark matter properties are concerned, the key features of this product group example is that the $T_i$ fields are electroweak singlets with degenerate tree-level masses
\be
\label{eq:productmass}
m_i = \lambda_{\rm top} f c_v,
\ee
and all three are exactly stable.  The na\"{\i}ve annihilation and direct detection cross sections are determined only by the coupling to the physical Higgs
\be
\label{eq:producthiggscoupling}
\mathcal{L}_{\rm int} = -\lambda_{\rm top} \left( \frac{s_v}{\sqrt{2}} T T^c h_0 + \frac{c_v}{4f} TT^c h_0^2 + \mathcal{O}(h_0^3)\right).
\ee
In particular, the thermal relic abundance can be calculated from the diagrams $T \overline{T} \rightarrow h_0^* \rightarrow W^+ W^-/ZZ/t \bar{t}$ and $T \overline{T} \rightarrow h_0 h_0$.

More generally, in a typical UV completion, there is no reason not to expect operators of the form
\be
\label{eq:4fermion}
\mathcal{L}_{\rm 4\, fermion} =  \frac{\gamma}{f^2} (\overline{q} \bar{\sigma}^\mu q)(\overline{T} \bar{\sigma}_\mu T) + \cdots,
\ee
where $\gamma$ is some coefficient.  This operator could arise, for example, from integrating out a heavy resonance of the composite sector, and leads to an enhancement of the $T \overline{T} \rightarrow t \bar{t}$ and $T \overline{T} \rightarrow b \bar{b}$ annihilation channels.  We will leave $\gamma$ as a free parameter, though we do not expect the magnitude of $\gamma$ to be smaller than $1/(4\pi)^2$ (the na\"{\i}ve dimensional analysis (NDA) \cite{Manohar:1983md,Georgi:1986kr} expectation if $T$ and $T^c$ are mostly elementary) or larger than $1$ (the NDA expectation if $T$ and $T^c$ are mostly composite).

\subsection{Simple Group Example}
\label{sec:simplegroup}

In the above example, the dark top fields were all electroweak singlets.  In simple group models, $SU(2)_L$ and $SU(3)_C$ are linked in a nontrivial way, allowing for $SU(2)_L$ doublet dark tops.

Consider the symmetry structure
\begin{align}
F &= SU(3)_C \times SU(2)_L, \nn\\
G &= SU(6), \nn\\
H &= SU(5),
\end{align}
where we have again suppressed hypercharge.\footnote{If one extends $SU(6)$ to $U(6)$, then hypercharge is the $\diag(-2/3,-2/3,-2/3,1/2,1/2,0)$ generator.  We will show how hypercharge works in the simple group model explicitly in \Sec{sec:explicitads}.}  We can break $G \rightarrow H$ using a linear sigma model field $\Phi$ that transforms as a $\mathbf{6}$ of $SU(6)$:
\be
\label{eq:goldstone65}
\Phi = e^{i \Pi/f} \left(\begin{array}{c}0 \\0 \\0 \\0 \\0 \\f\end{array}\right), \qquad \Pi =  \left(\begin{array}{cccccc}0 & 0 & 0 & 0 & 0 & \phi_r \\0 & 0 & 0 & 0 & 0 & \phi_g \\0 & 0 & 0 & 0 & 0 & \phi_b \\0 & 0 & 0 & 0 & 0 & h_1 \\0 & 0 & 0 & 0& 0 & h_2 \\ \phi^\dagger_r & \phi^\dagger_g & \phi^\dagger_b & h_1^\dagger & h_2^\dagger &0 \end{array}\right),
\ee
where $\phi$ is a complex color triplet Goldstone, $h$ is the ordinary Higgs boson, and we have suppressed the singlet $\eta$ that lives in $\diag(1,1,1,1,1,-5)$ since it will be eaten by a heavy gauge boson in our explicit UV completions.

We take $Q$ to be a $\mathbf{\overline{21}}$ (symmetric tensor) of $SU(6)$ and $Q^c$ to be a $\mathbf{6}$ of $SU(6)$.  The Yukawa interactions are
\be
\label{eq:Yukawa65}
\mathcal{L}_{\rm Yukawa} = \sqrt{2} \lambda_{\rm top}Q^{ij} \Phi_i Q_j^c,
\ee
with the decomposition of $Q$ and $Q^c$:
\be
\label{eq:decompose65}
Q = \frac{1}{\sqrt{2}} \left(\begin{array}{ccccccc} 0 & 0 & 0 & q_{1r} & q_{2r} & 0  \\0 & 0 & 0 & q_{1g} & q_{2g} & 0  \\0 & 0 &  0& q_{1b} & q_{1b} & 0 \\q_{1r} & q_{1g} & q_{1b} & 0 & 0 &  T_A \\q_{2r} & q_{2g} & q_{2b} & 0 & 0& T_B \\0 & 0 & 0 & T_A & T_B & T_C \sqrt{2} \end{array}\right), \qquad Q^c = \left(\begin{array}{c}t^c_r \\t^c_g \\t^c_b \\T^c_A \\T^c_B \\T^c_C\end{array}\right).
\ee
The factor of $\sqrt{2}$ in front of the $T_C$ term comes from canonical normalization.

Expanding out \Eq{eq:Yukawa65}, we have
\be
\mathcal{L}_{\rm Yukawa} = \lambda_{\rm top} \left[q h t^c + (T_A, T_B) h T_C^c + (T_A T_A^c + T_B T_B^c + \sqrt{2}T_C T_C^c) \left(f - \frac{h^\dagger h}{2f}\right) + \cdots  \right].
\ee
which yields the Yukawa structure of \Eq{eq:yukawa} with the advertised $\kappa^{ij}$ matrix of \Eq{eq:kappaexamplealt}. In \App{app:symmetries}, we show how to use symmetries to understand why the Higgs mass is protected.

The spectrum in the dark top sector is now quite interesting. The mass eigenstates are
\be
\label{eq:simplemass}
m_1 = \lambda_{\rm top} f c_v \left[1 - \frac{v^2}{2 f^2} + \cdots \right], \quad m_2 = \lambda_{\rm top} f c_v , \quad m_3 = \lambda_{\rm top} f c_v \left[ \sqrt{2} + \frac{v^2}{\sqrt{2} f^2} \cdots \right],
\ee
where the terms in brackets are the positive solutions to $x^2 + 2/x^2 = 3 + s_v^2/c_v^2$.  The state $m_1$ is mostly $SU(2)_L$ doublet and the state $m_3$ is mostly $SU(2)_L$ singlet but they mix because of electroweak symmetry breaking.  The lightest state $m_1$ is the only exactly stable state.

It is well-known that direct detection experiments place strong bounds on $SU(2)_L$ doublet dark matter from $Z$ exchange.  In order to suppress this interaction, we can use the method of inelastic dark matter \cite{TuckerSmith:2001hy}.  After electroweak symmetry breaking, the coupling
\be
\mathcal{L}_{\rm split} = \frac{\Phi^\dagger Q^c \Phi^\dagger Q^c}{M_{\rm split}}
\ee
gives a Majorana mass to a linear combination of $T_A^c$ and $T_C^c$, which splits the $Z$ interaction eigenstates of the lightest dark top (which is mostly $T_A$).  With a mass splitting above 1 MeV, direct detection through $Z$ exchange will no longer be kinematically allowed.

Just like in the product group example, one expects four-fermion operators like those in \Eq{eq:4fermion}, and we will include those in the dark matter phenomenology.  At minimum, after integrating out the PGB $\phi$, one expects a contribution to the four-fermion operator that gives an effective $\gamma_{\rm eff} \sim \lambda_{\rm top}^2 (f/m_\phi)^2$.  Because the operator generated by $\phi$ is slightly different from \Eq{eq:4fermion} and because we do not expect to be able to arbitrarily decouple $\phi$, we will simply include the $\phi$ couplings when studying the dark matter properties and choose a fiducial mass $m_\phi = 2 \lambda_{\rm top} f$.

While the couplings of the dark top to the Higgs are subdominant for determining the thermal relic abundance, they are the dominant contribution to the direct detection cross section once $Z$ exchange is kinematically forbidden.   For the lightest mass eigenstate $T_1$, the linear Higgs coupling is
\be
\label{eq:simplehiggscoupling}
\mathcal{L} = - \lambda_{\rm top} \left(\frac{\sqrt{2}v}{f} \left(1 - \frac{5 v^2}{3 f^2} + \cdots  \right) h_0 T_1 T_1^c \right).
\ee
Note that this coupling is approximately a factor of $2$ larger than the coupling in \Eq{eq:producthiggscoupling}.

Since the color triplet PGB $\phi$ is odd under the same $\mathbf{Z}_2$ symmetry as the dark top, one worries that $\phi$ would be the lightest $\mathbf{Z}_2$-odd particle.  But unlike the Higgs potential, the potential $V(\phi)$ is quadratically sensitive to the cutoff.  Using the Coleman-Weinberg potential
\be
\label{eq:phimass65}
m_\phi^2 \simeq \frac{1}{8 \pi^2} 2 \lambda_{\rm top}^2 \Lambda_{\rm TC}^2,
\ee
so for $\Lambda_{\rm TC} \simeq 4\pi f$, $\phi$ is heavier than the dark top.  In realistic composite models, $\Lambda_{\rm TC}$ is not usually that large, so one needs to work harder to make sure that $\phi$ is not the lightest $T$-odd particle.  Because $\phi$ carries color, $m_{\phi}$ gets a quadratically sensitive mass contribution from gluons, which allows smaller values of $\Lambda_{\rm TC}$ while still keeping $\phi$ heavier than the dark top.  To achieve much lower values of $\Lambda_{\rm TC}$ would require additional contributions to $m_\phi$, for example by introducing $G$-violating technifermion masses.

\subsection{Maximal Mixing?}
\label{sec:maxmix}

We saw that the product group example had pure $SU(2)_L$ singlet dark matter and the simple group example had nearly pure $SU(2)_L$ doublet dark matter.  Generically, one expects hybrid doublet/singlet dark matter to have ``nicer'' dark matter properties \cite{ArkaniHamed:2006mb}, meaning that the dark matter relic abundance is determined by thermal freezeout alone and the dark matter particle has some measurable cross section at the LHC. Can this ``well-tempered'' dark matter scenario occur in dark top models?

The answer is yes, but such a situation is disfavored for theoretical reasons.  In the $SU(6)_C \times SU(3)_L$ product group example, the standard model $SU(2)_L$ gauge group could be the diagonal combination of a $SU(2)$ subgroup of $SU(3)_L$ and a $SU(2)$ subgroup of $SU(6)_C$.  At tree level, the three $T_i$ fields will be arranged into an exactly degenerate doublet $T_d, T_d^c$ and singlet $T_s,T_s^c$.  Then one can add by hand a mixing term like
\be
\label{eq:mixingterm}
\mathcal{L}_{\rm mix} = \lambda_{\rm mix} \left(T_d h T_s^c + T_s h^\dagger T_d^c \right)
\ee
in the low energy effective theory. As long as $\lambda_{\rm mix}$ is not much bigger than $\lambda_{\rm bottom}$, then the quadratic divergence introduced by this mixing is no worse than the quadratic divergence from the non-top standard model fields.  Because the doublet and singlet were originally degenerate because of symmetries, \Eq{eq:mixingterm} will yield maximal mixing.

However, \Eq{eq:mixingterm} is immediately suspect because it violates the original $SU(6)_C \times SU(3)_L$ symmetry by hand, so it is not clear what kind of UV completion would yield an effective mixing term like this.  We emphasize that the problem is not just that this is a $G$-violating interaction---after all, the gauge couplings and Yukawa couplings already violate $G$---but that this interaction cannot even be lifted into a $G$-preserving interaction through the introduction of new fields.   To generate \Eq{eq:mixingterm} would require the UV completion to have additional sources of $G$ violation that only allow $\mathcal{L}_{\rm mix}$ and do not otherwise spoil the stability of the Higgs potential.

One might try to augment the original Yukawa coupling from \Eq{eq:yukawa} in some way to yield $\lambda_{\rm mix} = \lambda_{\rm top}$.  That is, the same term that generated the top Yukawa coupling could also generate maximal dark matter mixing.  Unfortunately, in order have $\lambda_{\rm mix} = \lambda_{\rm top}$ while still canceling the top contribution to the Higgs potential, one would need more dark top fields.  The reason is that \Eq{eq:mixingterm} generates the same kind of quadratic divergence as the top Yukawa itself, so doublet/singlet maximal mixing effectively yields $N_D < 3$.  In order to restore $N_D = 3$ one needs more dark top fields, but if those new fields also contribute the relic density, then the point of maximally mixing is lost.  Because the goal of the dark top was to have a dark matter particle that also regulates the Higgs potential, we disfavor a maximal mixing scenario.

\section{UV Completions}
\label{sec:uvcomplete}

While the low energy interactions of the dark top are sufficient to understand the dark matter properties, it is important to demonstrate that UV completions do exist.   As we already mentioned in \Sec{subsec:uvconsiderations}, in using the Coleman-Weinberg potential to show the cancellation of the top and dark top contributions to the Higgs radiative potential, we were assuming a uniform cutoff $\Lambda$ for all of the fermion fields.  For the analysis of \Sec{sec:darktop} to be valid, we have to show that such a uniform cutoff is achievable in practice.

Another reason to understand UV completions is that the dominant production mechanisms for new particles at the LHC are often pair- or associated-production of new heavy colored particles, so we want to understand the likely spectrum of colored particles in a dark top scenario.   \Eq{eq:yukawa} contains no new colored particles, but any concrete realization of the dark top Yukawa sector is likely (though not required) to invoke new colored particles.  We already saw the need for the color triplet PGB $\phi$ in the simple group example, but in specific UV completions there are additional types of colored particles possible in both the simple and product group examples.

\subsection{Tensions with the UV}
\label{sec:tension}

In \Sec{subsec:uvconsiderations}, we argued that simply having $N_D^{\rm eff} = 3$ was insufficient to guarantee that the top loop is properly regulated; one also needs a UV completion whose effective cutoff matrix is diagonal.  To emphasize the importance of having a valid UV completion it is instructive to write down a model that has the correct low energy field content to be a dark top, but cannot be UV completed in any straightforward way without destabilizing the Higgs potential.

To anticipate the result, we note that in the decomposition of the fermion fields in \Eq{eq:decompose63} and \Eq{eq:decompose65}, we zeroed out components in only one of $Q$ and $Q^c$ but not both.  For typical UV completions, there is an effective UV regulator scale $\Lambda_Q$ and a different UV scale $\Lambda_{Q^c}$, and there is no reason for those scales to be the same.  As long as only one of $Q$ or $Q^c$ violates the symmetries that protect the Higgs, then only one scale $\Lambda$ enters the radiative potential for the Higgs.  However, if we get the right low energy field content by zeroing out components of both $Q$ and $Q^c$, then the radiative potential will be sensitive to both $\Lambda_Q$ and $\Lambda_{Q^c}$, and for $\Lambda_Q \not= \Lambda_{Q^c}$ there will be large corrections to the radiative Higgs potential.

To get \Eq{eq:yukawa} at low energies in an inviable model, consider the symmetry breaking pattern
\begin{align}
F &= SU(3)_C \times SU(2)_L, \nn\\
G &= SU(7), \nn\\
H &= SU(6),
\end{align}
where we are again ignoring hypercharge.  The breaking of $G \rightarrow H$ can come from a linear sigma model field $\Phi$ that transforms as a $\mathbf{7}$ of $SU(7)$.  We can write $\Phi$ as
\be
\Phi = e^{i \Pi/f} \left(\begin{array}{c}0 \\0 \\0 \\0 \\0 \\0 \\f\end{array}\right), \qquad \Pi =  \left(\begin{array}{ccccccc}0 & 0 & 0 & 0 & 0 & 0 & \phi_r \\0 & 0 & 0 & 0 & 0 & 0 & \phi_g \\0 & 0 & 0 & 0 & 0 & 0 & \phi_b \\0 & 0 & 0 & 0 & 0 & 0 & h_1 \\0 & 0 & 0 & 0 & 0 & 0 & h_2 \\0 & 0 & 0 & 0 & 0 & 0 & \phi' \\ \phi^\dagger_r & \phi^\dagger_g & \phi^\dagger_b & h_1^\dagger & h_2^\dagger & \phi'^\dagger & 0 \end{array}\right),
\ee
where $\phi$ is a color triplet Goldstone, $h$ is the ordinary Higgs boson, $\phi'$ is a complex singlet, and we have suppressed the real singlet $\eta$ that lives on the diagonal.

Taking $Q$ to be a $\mathbf{\overline{21}}$ (anti-symmetric tensor) of $SU(7)$ and $Q^c$ to be a $\mathbf{7}$ of $SU(7)$, we can write the top Yukawa as
\be
\label{eq:Yukawa7}
\mathcal{L}_{\rm Yukawa} =  \sqrt{2} \lambda_{\rm top} Q^{ij} \Phi_i Q^c_j.
\ee
The decomposition of $Q$ and $Q^c$ is
\be
\label{eq:decompose7}
Q = \frac{1}{\sqrt{2}}\left(\begin{array}{ccccccc} 0& 0 & 0 & -q_{1r} & -q_{2r} & 0 & 0 \\0 & 0 & 0 & -q_{1g} & -q_{2g} & 0 & 0 \\0 & 0 & 0 & -q_{1b} & -q_{2b} & 0 & 0 \\q_{1r} & q_{1g} & q_{1b} & 0 & 0 & 0 & -T_A \\q_{2r} & q_{2g} & q_{2b} & 0 & 0 & 0& -T_B\\0 & 0 & 0 & 0 & 0 & 0 & -T_C \\0 & 0 & 0 & T_A & T_B & T_C & 0\end{array}\right), \qquad Q^c = \left(\begin{array}{c}t^c_r \\t^c_g \\t^c_b \\T^c_A \\T^c_B \\T^c_C \\ T^c_D\end{array}\right),
\ee
and we see that \Eq{eq:Yukawa7} does indeed yield \Eq{eq:yukawa} with the $\kappa_{ij}$ matrix of \Eq{eq:kappaexample} if we take $T^c_D \rightarrow 0$.

However, the precise way in which the components of $Q$ and $Q^c$ are zeroed out affects the cancellation of the top divergence.  Imagine decoupling $T^c_D$ by adding a vector-like partner $T_D$.  Then compared to \Eq{eq:yukawa}, the Lagrangian has the extra terms
\be
\mathcal{L}_{\rm extra} =  \lambda_{\rm top} (T_A h^\dagger_1 + T_B h^\dagger_2 + m_D T_D) T^c_D.
\ee
For any non-infinite value of $m_D$, there will be mixing between $T_D$ and $(T_A h^\dagger_1 + T_B h^\dagger_2)$.  This mixing will spoil the cancellation in \Eq{eq:trM2cancel}, because it would yield an effective $N_D^{\rm eff} < 3$.  So while this $SU(7)$ model has the right low energy field content, the radiative potential for the Higgs is sensitive to UV details of the model.

If there were a way to decouple the zeroed out fields in $Q$ and $Q^c$ at exactly the same rate, then we would find that the mixing between the low energy fields and the decoupling vector-like partners would still yield $N_D^{\rm eff} = 3$.  The reason this is unlikely to happen is that $Q$ and $Q^c$ have different quantum numbers, so it is hard to see how a UV completion could enforce a common decoupling mass scale.\footnote{You could achieve a common decoupling scale if $Q$ and $Q^c$ were part of a single multiplet $\tilde{Q}$ in a real representation of $G$.  This would allow the Yukawa coupling $\tilde{Q}\tilde{Q}\Sigma$.}  For the models in \Secs{sec:productgroup}{sec:simplegroup}, the zeroed out fields are only in $Q$, so as long as the UV completion respects the $G$ global symmetry, the decoupling of the zeroed out components can be uniform and $N_D^{\rm eff}$ will indeed equal $3$.  The UV completions we present next respect $G$ by construction.

\subsection{Explicit AdS$_5$ Construction}
\label{sec:explicitads}

\begin{figure}
\begin{center}
\includegraphics[scale=0.8]{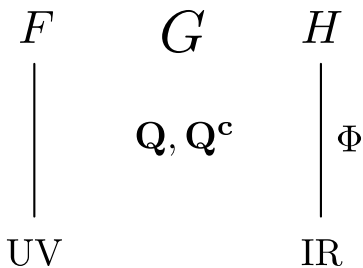}
\end{center}
\caption{An AdS$_5$ construction that is holographically dual to a strongly coupled 4D field theory with a $G$ global symmetry.  In the 4D theory, the global symmetry $G$ is spontaneously broken to $H$, which is represented in AdS$_5$ by reducing the bulk $G$ gauge symmetry to $H$ on the IR brane.  In the 4D theory, an $F$ subgroup of $G$ is explicitly gauged, which is represented in AdS$_5$ by reducing the bulk $G$ gauge symmetry to $F$ on the UV brane.  For calculational convenience, we choose to implement the IR brane symmetry breaking through a brane-localized linear sigma field $\Phi$ and the UV brane symmetry breaking through Dirichlet boundary conditions. The top and dark top are contained in the bulk fermions $\mathbf{Q}$ and $\mathbf{Q^c}$. \label{fig:adsexample}}
\end{figure}

A natural UV context for a dark top model is a composite scenario.  The models in \Secs{sec:productgroup}{sec:simplegroup} naturally fit into the framework of strong dynamics with an approximate $G$ global symmetry that is spontaneously broken to $H$ at some scale $\Lambda_{\rm TC}$.  A subgroup $F \subset G$ is gauged to yield the standard model gauge group in $F \cap H$ at low energies.  As shown in \Fig{fig:adsexample}, such composite models have approximate AdS$_5$ duals with a bulk $G$ gauge symmetry that is reduced to $F$ on the UV brane and $H$ on the IR brane \cite{Maldacena:1997re, Gubser:1998bc, Witten:1998qj, Randall:1999ee, ArkaniHamed:2000ds, Rattazzi:2000hs,Contino:2003ve}.

For concreteness, take the simple group example from \Sec{sec:simplegroup}.  This model can come from a slice of AdS$_5$ with a bulk $SU(6)$ gauge symmetry.  For calculational purposes, it is convenient to localize a linear sigma model $\Phi$ on the IR brane to break $SU(6) \rightarrow SU(5)$.\footnote{In general, the PGBs in $\Phi$ will mix with the $(-,+)$ Kaluza-Klein (KK) gauge bosons, inducing wavefunction renormalization for the Higgs field.  We will ignore this effect in our discussion, but it is important to get the right numerics.}  The bulk gauge symmetry is reduced to the standard model gauge group on the UV brane using boundary conditions.  The fermions $Q$ and $Q^c$ live in the bulk as part of 5D Dirac multiplets $\mathbf{Q}$ and $\mathbf{Q^c}$
\be
\mathbf{Q} = \left(\begin{array}{c} Q\\ \overline{Q'} \end{array}  \right), \qquad \mathbf{Q^c} = \left(\begin{array}{c} Q'^c \\ \overline{Q^c} \end{array}  \right),
\ee
where we have chosen $Q$ to be in the upper component of $\mathbf{Q}$ and $Q^c$ to be in the lower component of $\mathbf{Q^c}$ in order to use the results from \cite{Thaler:2005en}.   We take UV/IR $(-,+)$ boundary conditions for the zeroed-out entries from \Eq{eq:decompose65} and $(+,+)$ boundary conditions on the remaining components.\footnote{Recall that the boundary conditions on $Q'$ are the reverse of the boundary conditions on $Q$ \cite{Henningson:1998cd,Mueck:1998iz}.}  The coupling $\hat{\lambda} Q \Phi Q^c$ is localized on the IR brane, and $\hat{\lambda}$ determines the top Yukawa coupling once wave function overlaps are taken into account.

To get the correct hypercharges in the AdS$_5$ construction, extra $U(1)$ gauge fields are needed in the bulk \cite{Contino:2003ve,Thaler:2005en}.  By choosing appropriate UV and IR boundary conditions, one can make sure that the only $U(1)$ zero mode is hypercharge and that there are no $\mathbf{Z}_2$-odd spin-0 or spin-1 states lighter than the dark top (otherwise the dark top would not be the dark matter candidate).  In the simple group model, we can add an extra bulk $U(1)_X$ gauge field that is unbroken on the IR brane.  The full bulk $SU(6) \times U(1)_X$ gauge symmetry is reduced to
\be
SU(3)_C \times SU(2)_L \times U(1)_Y \times U(1)_\eta
\ee
on the UV brane using boundary conditions.  It is convenient to talk about $SU(6) \times U(1)_X$ as if it were a single $U(6)$ group, in which case the (unnormalized) $U(1)_Y$ and $U(1)_\eta$ generators are\footnote{The choice of $T_\eta$ is ambiguous.  We have chosen a generator orthogonal to $T_Y$ in order to minimize kinetic mixing.}
\be
T_Y = \diag(-2/3, -2/3, -2/3, 1/2, 1/2, 0), \qquad T_\eta = \diag(1/6,1/6,1/6,1/3,1/3,-7/6),
\ee
and the $U(6)$ transformation properties of $\mathbf{Q}$, $\mathbf{Q^c}$, and $\Phi$ are inherited from their $SU(6)$ representation.  The $U(1)_\eta$ gauge boson eats the $\eta$ Goldstone mentioned below \Eq{eq:goldstone65}, and in order for this field to be heavier than the dark top, the bulk $SU(6)$ gauge coupling must be large.   UV brane kinetic terms can then be used to recover the standard model gauge couplings for $SU(3)_C \times SU(2)_L \times U(1)_Y$.

The simplest way to add the other standard model fermions to this construction is to introduce additional quark and lepton multiplets in the bulk of AdS but impose UV boundary conditions such that the only light fields are the standard model fermions.  By using sufficiently exotic $U(6)$ representations, one can always achieve the correct Yukawa couplings to the Higgs on the IR brane.  To get smaller $U(6)$ representations, one would need to introduce separate up- and down-type $\Phi$ fields.  We will postpone a detailed discussion of flavor to future work, but like other composite Higgs models, the physical Yukawa couplings and mixings will be determined by a combination of the IR brane Higgs coupling and the fermion wavefunction overlap with the IR brane \cite{Grossman:1999ra,Gherghetta:2000qt,Huber:2000ie,Huber:2003tu}.

Because we are invoking a $\mathbf{Z}_2$ symmetry to keep the dark top stable, one worries whether this discrete symmetry is anomalous in the particular UV completion we have chosen.  In particular, Chern-Simons terms are usually needed in the bulk of AdS$_5$ in order to satisfy the AdS$_5$ anomaly matching conditions \cite{Callan:1984sa}.  Chern-Simons terms can violate $T$-parity in composite Higgs models, leading to the decay of the lightest $T$-odd particle \cite{Hill:2007nz,Hill:2007zv}.  In the case of the dark top, though, the dark matter candidate is a color singlet fermion with baryon number $1/3$, so the only way for it to decay to standard model fields is if both baryon number and the $\mathbf{Z}_2$ symmetry are violated.  Assuming baryon number violation is sufficiently weak, the dark top will have a very long lifetime even in the presence of bulk Chern-Simons terms.

\subsection{Radiative Higgs Potential}
\label{subsec:adshiggs}

We now can check that the top contribution to the Higgs potential exhibits the desired dark top cancellation in the AdS construction.  Because the Higgs is localized on the IR brane but the symmetries that protect the Higgs are broken only on the UV brane, the radiative Higgs potential is finite.  As shown in Ref.~\cite{Thaler:2005en}, the one-loop radiative corrections to the Goldstone potential from fermions can be written as
\be
V(\Phi) = -2\,  \tr \int \frac{p^3 dp}{8 \pi^2} \log\left(1 + p^2 \, \hat{G}(p) \cdot M \cdot \hat{G}^c(p) \cdot M^\dagger \right),
\ee
where $\hat{G}(p)$ ($\hat{G}^c(p)$) are rescaled IR boundary-to-boundary propagators for $Q$ ($Q^c$) and $M = \hat{\lambda} \Phi$ is the IR brane localized mass term for $Q$ and $Q^c$.  The rescaled propagators have dimension $(\mbox{mass})^{-2}$.  Calling the $(+,+)$ propagators $G$ and $G^c$ and the $(-,+)$ propagators $G_{\rm br}$ and $G^c_{\rm br}$,\footnote{There are no fields that use $G^c_{\rm br}$ for the $SU(6)/SU(5)$ model because all of the components of $Q^c$ have $(+,+)$ boundary conditions.  However, $G^c_{\rm br}$ is needed for the $SU(7)/SU(6)$ model.} it is straightforward to expand $V(\Phi)$ to extract the radiative correction to the Higgs mass.  Up to numeric prefactors, the Higgs mass correction in the $SU(6)/SU(5)$ simple group model is
\be
\label{eq:mhinads6}
\delta m_h^2 \propto \int \frac{p^3 dp}{8 \pi^2} \frac{\hat{\lambda}^4 p^4 \left(G - G_{\rm br}\right)^2 (G^c)^2}{\left(1 + \hat{\lambda}^2 p^2 G G^c \right) \left(1 + \hat{\lambda}^2 p^2 G_{\rm br} G^c \right)}.
\ee
The main thing to notice about this result is that $\delta m_h^2$ starts at order $\hat{\lambda}^4$.  Because the low energy quadratic divergence is proportional to $\tr |M|^2$ which is order $\hat{\lambda}^2$, we have shown that the low energy quadratic sensitive is removed and we are left with only a logarithmic sensitivity.  As expected, $\delta m_h^2 \rightarrow 0$ when $G_{\rm br} \rightarrow G$ since in that limit the Yukawa interaction respects all the symmetries that protect the Higgs mass.

It is instructive to contrast \Eq{eq:mhinads6} with the equivalent calculation for the $SU(7)/SU(6)$ model in \Sec{sec:tension}.  In that case,
\be
\label{eq:mhinads7}
\delta m_h^2 \propto \int \frac{p^3 dp}{8 \pi^2} \frac{\hat{\lambda}^2 p^2 \left(G G^c_{\rm br} - G_{\rm br} G^c \right) + \mathcal{O}(\hat{\lambda}^4)}{\left(1 + \hat{\lambda}^2 p^2 G G^c \right) \left(1 + \hat{\lambda}^2 p^2 G_{\rm br} G^c \right)}.
\ee
We see that this result starts at order $\hat{\lambda}^2$, which signals quadratic sensitivity from the low energy point of view.  In AdS$_5$ implementations of little Higgs theories, $\delta m_h^2$ also starts at order $\hat{\lambda}^2$, but in that case the numerator scales like $(G - G_{\rm br})(G^c - G_{\rm br}^c)$ indicating the presence of collective symmetry breaking \cite{Thaler:2005en}.  While $(G G^c_{\rm br} - G_{\rm br} G^c)$ goes to zero in the limit that both $G_{\rm br} \rightarrow G$ and $G_{\rm br}^c \rightarrow G^c$, it does not go to zero in either limit alone, and this means that \Eq{eq:mhinads7} is numerically large.  In \App{sec:deconstructed}, we do an equivalent deconstructed version of the AdS$_5$ calculation to gain more insight for why the cancellation works in the $SU(6)/SU(5)$ model but not in the $SU(7)/SU(6)$ model.

We have only calculated the contribution to the Higgs potential from top loops; there are also potentially large contributions from $SU(2)_L$ gauge loops.  From a low energy point of view, these are quadratically sensitive to the cutoff, but because all of the symmetries that protect the Higgs are restored in the bulk of AdS, the effective cutoff is the mass of the first KK mode which we denote by $\Lambda_{\rm TC}$.  One can do the AdS calculation explicitly to verify that the electroweak gauge loop is regulated at the KK mass scale, therefore the estimate in \Eq{eq;gaugeovertop} holds for the AdS construction.

Because the dark top is color neutral, there are potentially large two-loop contributions to the Higgs potential involving gluons and tops that have no low energy counterpart for dark tops.  This will contribute a two-loop quadratic divergence to the Higgs mass.  At higher energies, though, there are additional ``dark gluons'' (\ie KK gauge bosons) that couple to the dark tops, and because the symmetries that protect the Higgs are restored in the AdS bulk, we know that the gluon/top loops must cancel against dark gluons/dark top loops at sufficiently high energies.  Though we have not done an explicit calculation, we expect that the two-loop quadratic divergence is cancelled at the mass of the first KK mode and is therefore sufficiently small not to destabilize the Higgs potential.  This expectation is guided by the fact that the only operators that can differentiate between the $SU(3)_C$ quantum numbers of the top and dark top live on the UV brane, so the two-loop result should be suppressed by the effective size of the AdS space, which is governed by the first KK mass.

\subsection{UV Field Content}
\label{subsec:uvfield}

In addition to the low energy fields from \Sec{sec:models}, there are additional KK modes in the AdS$_5$ construction that may be produced at colliders.  This feature is not unique to AdS, of course; if there is a strongly interacting sector with a global symmetry $G$ that breaks to $H$ at some scale $\Lambda_{\rm TC}$, there are typically effective low energy fields with masses around $\Lambda_{\rm TC}$ that transform (nonlinearly) under $H$ \cite{Coleman:1969sm,Callan:1969sn}.  For example, if the top is a composite field then $H$ must contain $SU(3)_C$, which typically yields a tower of spin-1 color adjoint (\ie KK gluon) modes.

The main novelty of the dark top KK spectrum compared to other composite Higgs models is that $H$ can have non-trivial transformation properties under $SU(3)_C$, allowing for heavy spin-1 resonances that transform in the fundamental of $SU(3)_C$ and not just in the adjoint.  The reason is that the top and the dark top are related by the $G$ global symmetry, so there are generators in $H$ that rotate the top into the dark top.  For this rotation to make sense, these generators must be in the $\mathbf{3}$ of $SU(3)_C$, allowing for spin-1 fields that couple the top to the dark top.  The only exception is if the top and dark top are connected only by a discrete symmetry, in which case there need not be such spin-1 resonances.  If such fields exist, they must be odd under the $\mathbf{Z}_2$ symmetry that stabilizes the dark top.

In the product group example, the $SU(6)_C$ global symmetry is unbroken at low energies, but only an $SU(3)_C$ subgroup is gauged.  Therefore, there are three color triplet spin-1 bosons $X_i$ that couple $t^c$ to $T_i^c$, namely the $(-,+)$ KK gauge bosons.  Since the $T_i^c$ are neutral under hypercharge, the $X_i$ must have the opposite hypercharge compared to $t^c$, and this hypercharge assignment forbids $X_i$ from coupling $q$ to $T_i$.   In the simple group example, the $SU(6)$ global symmetry is broken to $SU(5)$ at low energies, and an $SU(3)_C \times SU(2)_L$ subgroup is gauged.  Therefore, there is a $(\mathbf{3},\mathbf{2})_{1/6}$ $X_d$ gauge boson that couples $t^c$ to $(T^c_A,T^c_B)$ and a $(\mathbf{3},\mathbf{1})_{2/3}$ $X_s$ gauge boson that couples $t^c$ to $T^c_C$ and $q$ to $(T_A,T_B)$.

For any of these colored $X$ gauge bosons to be kinematically accessible at the LHC, $\Lambda_{\rm TC}$ has to be around $1 \TeV$, especially since the interesting $X$ fields are $\mathbf{Z}_2$-odd and must be pair-produced.  However, a low value of $\Lambda_{\rm TC}$ does introduce tension with the assumed hierarchy $v \ll f \ll \Lambda_{\rm TC}$.  The decay constant scales like \cite{Contino:2003ve}
\be
f \sim \frac{\Lambda_{\rm TC}}{g_\rho},
\ee
where we are defining $\Lambda_{\rm TC}$ to be the mass of the first KK modes, and $g_\rho$ is the perturbative AdS expansion parameter which has the holographic interpretation of ${4 \pi}/\sqrt{N}$ in a large $N$ CFT \cite{ArkaniHamed:2000ds}.  Large values of $g_\rho$ imply a breakdown of perturbation theory on the AdS side, so perturbativity arguments favor smaller ratios of $\Lambda_{\rm TC}/f$.   In addition, if we do not want large corrections to the Higgs potential from gauge boson loops, then \Eq{eq;gaugeovertop} prefers small $\Lambda_{\rm TC}/f$.  On the other hand, in \Eq{eq:phimass65} we needed a large ratio of $\Lambda_{\rm TC}/f$ to explain why the $\phi$ Goldstone was heavy enough.   Also, larger values of $\Lambda_{\rm TC}$ are better for reducing the contribution to precision electroweak observables from KK modes \cite{Agashe:2005dk}.

This tension between two different preferred limits for $\Lambda_{\rm TC}/f$ is irreducible.  In studying the dark top phenomenology, we will assume that there are moderate values of $g_\rho$ for which the AdS$_5$ description is reliable, the dark top is the correct dark matter candidate, fine-tuning is manageable, and precision electroweak tests are satisfied.  For interesting LHC signatures, one hopes that $\Lambda_{\rm TC}$ is still light enough that some of the $X$ fields are kinematically accessible, though we emphasize that in the simple group model, the color triplet $\phi$ Goldstone can still be produced even if $f \ll \Lambda_{\rm TC}$.

Finally, in addition to new colored $\mathbf{Z}_2$-odd spin-1 field, typical UV completions contain $\mathbf{Z}_2$-even spin-1 fields. These $\mathbf{Z}_2$-even fields are found in many composite Higgs scenarios, but their effect on the dark top sector can be particularly important, since they can generate effective four-fermion operators like those in \Eq{eq:4fermion} when they are integrated out.  In AdS constructions, the size of these four-fermion operators depend on the overlap of the fermion and spin-1 wave functions and are therefore highly model dependent.  In determining the dark matter properties of the dark top, we will leave the size of these four-fermion operators as a free parameter.

\section{Dark Matter Properties}
\label{sec:darkmatter}

We now turn to studying the dark matter properties of dark tops.  In this section, we will calculate the dark top annihilation cross section to determine the expected thermal relic abundance, as well as the current and future bounds placed on the dark top mass from direct detection experiments.  Both of these properties are dominantly determined by the quantum numbers of $T$, $T^c$ and by the mass matrix $\kappa^{ij}$, at least for UV completions which do not give large contributions to the four-fermion operators of \Eq{eq:4fermion}.  We will also illustrate how large the UV sensitive four-fermion operators need to be in order to have an appreciable effect on the annihilation cross section.

For each of the dark top models presented in \Sec{sec:models}, the preferred value of the dark top mass can be determined by requiring that the model reproduce the observed thermal relic abundance.  For this, we consider a model acceptable if it falls within the $2\sigma$ range of the five-year Wilkinson Microwave Anisotropy Probe (WMAP) results combined with cosmological and astrophysical data, $0.1075 \leq \Omega_{\rm DM} h^2 \leq 0.1211$ \cite{Komatsu:2008hk}.  One should keep in mind, however, that the actual dark top mass can be different from the preferred value if there are non-thermal production mechanisms or if the dark top does not compose all of the dark matter.  In the case of non-thermal production, there are constraints on the dark top mass from direct detection as well as indirect detection.  We will only consider direct detection constraints in this paper, though in the simple group model with a large annihilation cross section to $W$s, recent indirect detection studies \cite{Grajek:2008jb,Nagai:2008se} may also be relevant.

\subsection{Thermal Relic Abundance}

Using \Eqs{eq:yukawa}{eq:4fermion} along with standard thermal relic abundance arguments, we can determine the expected dark top relic abundance as a function of the decay constant $f$ for each of the models in \Sec{sec:models}.   We use the \texttt{micrOMEGAs 2.2} \cite{Belanger:2006is} package which facilitates the calculation of tree-level annihilation cross sections and corresponding relic abundances from \texttt{CompHEP} \cite{Boos:2004kh}/\texttt{CalcHEP} \cite{Pukhov:2004ca} model files.\footnote{\texttt{LanHEP 3.0} \cite{Semenov:2008jy} was also used to check the implementation of the new models.}  For comparison with other studies, it is convenient to translate the decay constant $f$ into the lightest dark top mass $m_{\rm dark}$ using \Eqs{eq:productmass}{eq:simplemass}.  Note that $f$ is equal to $m_{\rm dark}$ to leading order in $v/f$.     For each model, we will show what range of $m_{\rm dark}$ yields the preferred WMAP value of $\Omega_{\rm DM}$.

\begin{figure}
\begin{center}
\includegraphics[scale=0.62]{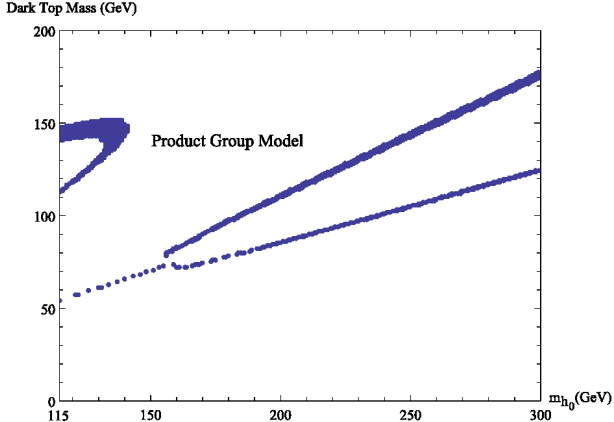}
\end{center}
\caption{The dark top masses $m_{\rm dark}$ that reproduce the $2\sigma$ WMAP range $0.1075 \leq \Omega_{\rm DM} h^2 \leq 0.1211$ as the physical Higgs mass $m_{h_0}$ is varied.  We only show the product group model from \Sec{sec:productgroup} since the simple group model from \Sec{sec:simplegroup} does not exhibit large sensitivity to $m_{h_0}$.  Over  a wide range of Higgs masses, the preferred product group dark top mass is less than $m_{\rm top}$, violating the assumed hierarchy $v \ll f$.  In \Fig{fig:gammaplot}, we show how UV-sensitive operators can help increase the preferred mass range.   An interesting feature of this plot is the bifurcation of the preferred mass range above $m_{h_0} = 160 \GeV$, coming from valid values of $m_{\rm dark}$ appearing on either side of the Higgs resonance region $2 m_{\rm dark} \sim m_{h_0}$.\label{fig:higgsplot}  }
\end{figure}

For the product group model, the dark top is an $SU(2)$ singlet and dominantly annihilates through the channels $T \overline{T} \rightarrow h_0 h_0$ and $T \overline{T} \rightarrow h_0^* \rightarrow WW/ZZ/t \bar{t}$ with the operator in \Eq{eq:producthiggscoupling}.  However, these cross sections are fairly small, and as shown in \Fig{fig:higgsplot}, the dark top is required to be quite light, with $m_{\rm dark} \sim v$.\footnote{With such large values of $v/f$, the couplings of the Higgs to the standard model gauge bosons and fermions are substantially different from an ordinary single Higgs doublet.  We have accounted for this effect in \texttt{micrOMEGAs} by using the $\Phi$ expansion in \Eq{eq:phiexpansioninh0}, and roughly speaking, the linear couplings of the Higgs to standard model fields is suppressed by a factor of $c_v$ compared to the standard model expectation.}  The reason for this is that, in the nonrelativistic expansion $\sigma v_{\rm rel} \approx a + b v_{\rm rel}^2$, these processes have $a = 0$.  The thermally averaged cross section is then suppressed by a factor of $\frac{1}{x_f} \equiv \frac{T_f}{m_{\rm dark}} \approx \frac{1}{20}$, where $T_f$ is the freezeout temperature.  While light dark tops are not ideal---the $v/f$ expansion breaks down, there may be large contributions to electroweak observables from physics at the scale $\Lambda_{\rm TC}$, and there are strong direct detection constraints---we will see that this scale can easily be raised to a more acceptable value by not completely decoupling the contributions from UV physics.

In contrast, for the simple group model, the dark top has a large $SU(2)$ doublet component and (co)annihilates very efficiently through the channels $T \overline{T'} \rightarrow WW/WZ/ZZ$ and $T \overline{T'} \rightarrow W^* \rightarrow \psi \overline{\psi'}$ where $\psi$ represents standard model fermions.  Because there is only a small splitting between the lightest mass eigenstates, coannihilation effects are important.  With such efficient annihilation, the preferred dark top mass is somewhat large, $980 \GeV < m_{\rm dark}< 1040 \GeV$, and potential contributions from UV physics only increase the preferred mass range further.  Because annihilations to Higgs and through Higgses are subdominant, the preferred mass range shows very little sensitivity to the physical Higgs mass.

\begin{figure}
\begin{center}
\includegraphics[scale=0.62]{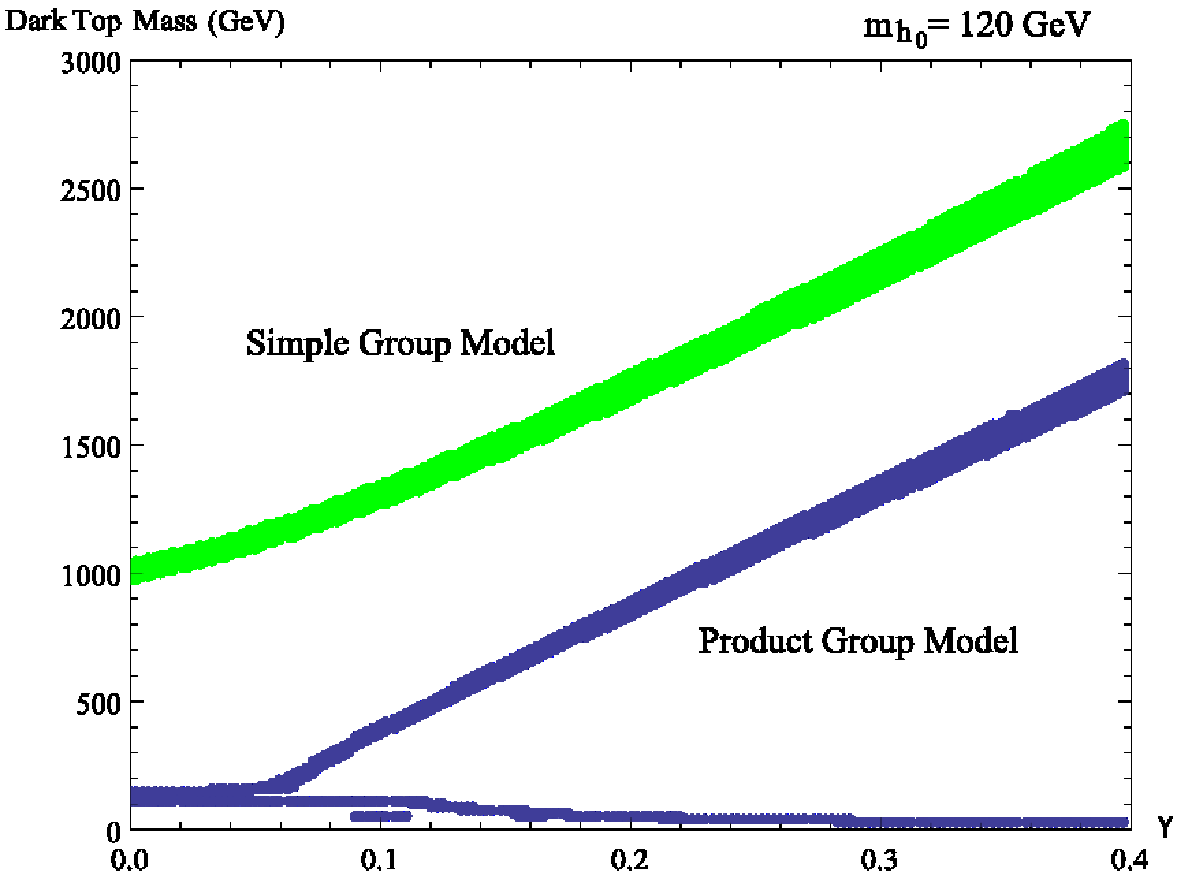}
$\quad\quad$
\includegraphics[scale=0.62]{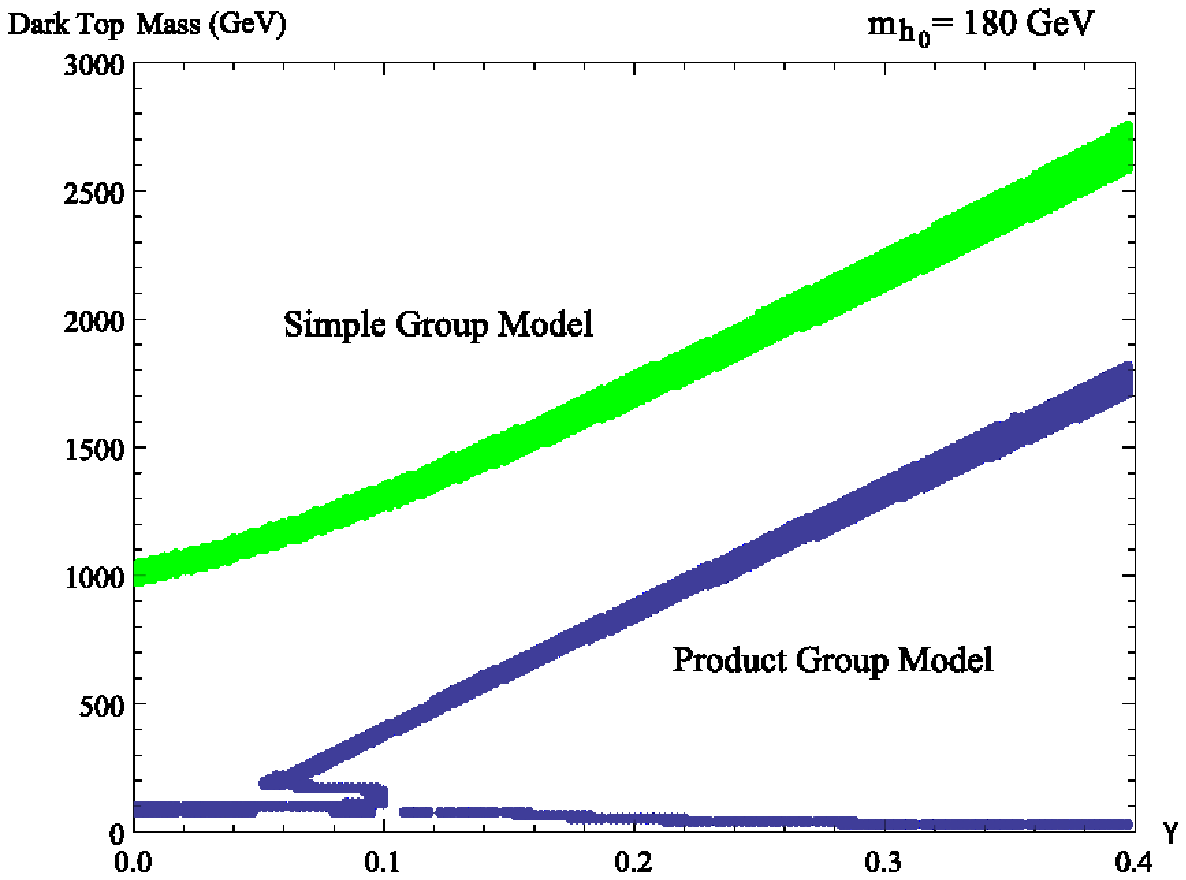}
\end{center}
\caption{The dark top masses $m_{\rm dark}$ that reproduce the $2\sigma$ WMAP range $0.1075 \leq \Omega_{\rm DM} h^2 \leq 0.1211$ as we vary the coefficient $\gamma$ of the four-fermion operator from \Eq{eq:4f}.  The left plot has a fixed physical Higgs mass $m_{h_0} = 120 \GeV$ and the right plot has $m_{h_0} = 180 \GeV$.   For $\gamma$ larger than $\sim .06$, the UV-sensitive operators dominate over the UV-insensitive annihilation mechanisms.   Note that the large $\gamma$ behavior is different for the two models because there are three species of stable dark top in the product group model but only one species of stable dark top in the simple group model.  In the simple group model, we have fixed $m_\phi = 2 \lambda_{\rm top} f$, and because $\phi$ gives an effective contribution to the four-fermion operators, this plot is not symmetric around $\gamma = 0$.  The existence of the lower branch in the product group model is due to being in a region of parameter space where annihilation to top quarks is no longer kinematically allowed, so the four-fermion operator only allows annihilation to bottom quarks.  Such a region badly violates the $v/f$ expansion and is theoretically disfavored.
\label{fig:gammaplot} }
\end{figure}

The effect of UV-sensitive operators are presented in \Fig{fig:gammaplot}.  We fix the Higgs mass at two fiducial values $m_{h_0} = 120 \GeV$ and $m_{h_0} = 180 \GeV$ and vary the coefficient $\gamma$ of one potential UV-sensitive four-fermion operator
\be
\label{eq:4f}
\mathcal{L}_{\rm 4\, fermion} =  \frac{\gamma}{f^2} (\overline{q} \bar{\sigma}^\mu q)(\overline{T} \bar{\sigma}_\mu T),
\ee
where $q$ contains both the left-handed top and bottom of the standard model, and $T$ contains the three dark top particles $T_A$, $T_B$, and $T_C$.   This operator enhances the annihilation cross section $T\overline{T} \rightarrow t\bar{t}/b\bar{b}$.  When the four-fermion coupling $\gamma$ is larger than $\sim .06$, the annihilation in both models is dominated by the UV-sensitive operators.  One thing to note about \Fig{fig:gammaplot} is that although the coupling $\gamma$ in both models is assumed to be the same, in the product group model there are effectively three species of dark matter particles, whereas in the simple group model there is only one, which explains why $m_{\rm dark}$ needs to be larger in the simple group model than the product group model to get the same relic density.

There are of course other four-fermion operators involving the dark top that may be generated by UV physics, for example
\be
\frac{\gamma'}{f^2} (\overline{t^c} \bar{\sigma}^\mu t^c)(\overline{T} \bar{\sigma}_\mu T),
\ee
and for operators like this that only involve right-handed tops, there is no enhanced annihilation to bottom quarks.  For heavy enough dark tops, the difference between the top mass and bottom mass can be ignored and the dominant effect can be absorbed into an effective $\gamma_{\rm eff} \simeq \gamma'/\sqrt{2}$.  We have also checked that the effect of all other four-fermion operators involving dark tops and third generation quarks can be dominantly described by a suitable choice of $\gamma_{\rm eff}$.

\subsection{Direct Detection Bounds}
Direct detection experiments designed to search for scattering between dark matter and heavy nuclei give strong constraints on the allowed dark top parameter space.  These experiments constrain the interaction cross section as a function of the mass of the dark matter, given an assumed local dark matter mass density $\rho_0$ and velocity $v_0$.  We will make the standard assumptions of $\rho_0 \simeq 0.3 \mbox{ GeV/cm$^3$}$ and $v_0 \simeq 230 \pm 20 ~\mathrm{km/s}$ \cite{Kamionkowski:1997xg,Amsler:2008zz}, and calculate cross sections using \texttt{micrOMEGAs 2.2} \cite{Belanger:2006is}.

\begin{figure}
\begin{center}
\includegraphics[scale=0.75]{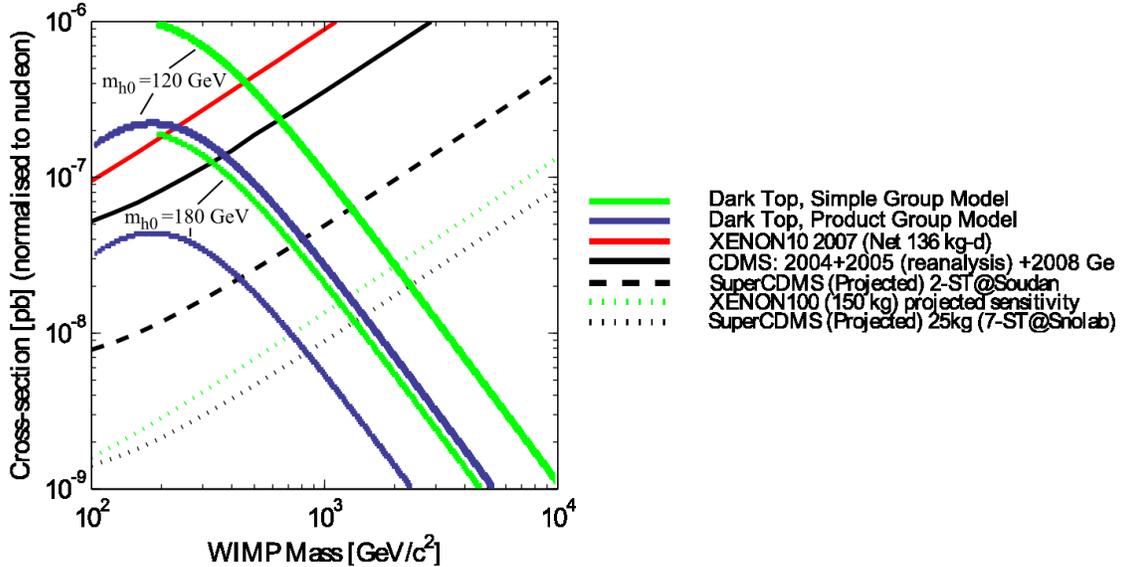}
\end{center}
\caption{The dark top direct detection cross section per nucleon versus the dark top mass, as compared with the current and projected future bounds from the CDMS \cite{Ahmed:2008eu} and XENON \cite{Angle:2007uj} experiments.  Shown are the product group model from \Sec{sec:productgroup} and the simple group model from \Sec{sec:simplegroup}, with fixed physical Higgs masses $m_{h_0} = 120 \GeV$ and $m_{h_0} = 180 \GeV$.  The limit plots were generated using \texttt{DMTools} \cite{DMTools}. \label{fig:directplot}}
\end{figure}

In dark top models, the direct detection cross sections are essentially determined by the symmetries of the model, the top Yukawa coupling, and the Higgs mass.  UV-sensitive four-fermion operators do not change the cross sections appreciably.  In the product group model, the dark top scatters off protons and neutrons primarily through $t$-channel Higgs exchange.  Note that this only gives a contribution to the spin-independent cross section.  As shown in \Fig{fig:directplot}, the strongest bounds on the dark top mass currently come from the CDMS \cite{Ahmed:2008eu} and XENON \cite{Angle:2007uj} experiments, and we find that these bounds rule out dark top masses below $\sim 350 \GeV$ for $m_{h_0} = 120 \GeV$ but impose no constraints for $m_{h_0} = 180 \GeV$.   In addition, future upgrades to these experiments should be able to probe a significantly larger portion of parameter space.

In contrast, in the simple group model the dark top carries non-zero hypercharge, and na\"{\i}vely has a large coupling to the $Z$.  As is well known, for weak scale masses this leads to elastic cross sections with nuclei that are several orders of magnitude above present bounds, and is strongly ruled out \cite{Goodman:1984dc,Cirelli:2005uq,Essig:2007az}.  As discussed in \Sec{sec:simplegroup}, if we give a small Majorana mass to the dark top, this splits the Majorana mass eigenstates that couple to the $Z$ and can make scattering through $Z$ exchange kinematically disallowed.  In particular, we consider here adding a Majorana mass $\delta$ in the range $1 \MeV < \delta < 100 \MeV$, corresponding to introducing the coupling
\be
\mathcal{L}_{\rm split} = \frac{\Phi^\dagger Q^c \Phi^\dagger Q^c}{M_{\rm split}},
\ee
and choosing $M_{\rm split} \sim (\mathrm{few} - 10^2) ~f$.\footnote{Although it na\"{\i}vely looks like this gives a Majorana mass $\delta \sim \frac{v^2}{M_{\rm split}}$ to $T^c$, after diagonalizing the Dirac masses the leading contribution actually comes as $\delta \sim \frac{v^6}{f^4 M_{\rm split}}$.}  With a splitting of this size, scattering off of nuclei again happens predominantly through Higgs exchange.  However, as one can see by comparing \Eqs{eq:producthiggscoupling}{eq:simplehiggscoupling}, the linear Higgs coupling in the simple group model is larger by a factor of $2$ compared to the product group model.  Because of this, the bounds in this model are even stronger, with CDMS ruling out masses below $\sim 700 \GeV$ for $m_{h_0} = 120 \GeV$ and below $\sim 300 \GeV$ for $m_{h_0} = 180 \GeV$.  In addition, future upgrades should be able to probe the preferred region $m_{\rm dark} \sim 1 \TeV$ in this model.

\section{LHC Signatures}
\label{sec:lhc}

Just like for neutralinos in supersymmetric models, direct production of the dark tops is a challenging discovery strategy \cite{atlasTDR,cmsTDR}.  A more promising discovery mode is to produce dark tops in cascade decays from new colored particles.  We have seen that in order for there to be a cancellation between loops of tops and dark tops, they must be in a common multiplet of some global symmetry.  In the most pessimistic scenario, the only link between the two is a discrete symmetry, in which case there need not be new colored particles.  But if the top and dark top are linked by a continuous symmetry, then there must be a symmetry generator that transforms like a triplet of $SU(3)_C$, and it is likely that there is some field that transforms according to that generator.

From the discussion in \Secs{sec:simplegroup}{subsec:uvfield}, the two likely candidates for a new color triplet field $X$ are pseudo-Goldstone bosons (which occurs in our simple group model) and heavy spin-1 resonances (which occurs in both our product and simple group models).  For spin-0 $X$, the coupling to the dark top are determined mainly by the top Yukawa $\lambda_{\rm top}$ through operators like
\be
\mathcal{L}_{\rm spin-0} \simeq \lambda_{\rm top} T X t^c .
\ee
For spin-1 $X_\mu$, the coupling to the dark top are determined mainly by gauge or gauge-like interactions such as
\be
\mathcal{L}_{\rm spin-1} \simeq g_{\rm TC} \overline{t^c} \bar{\sigma}^\mu T X_\mu.
\ee
In either case, the $X$ is odd under the $\mathbf{Z}_2$ symmetry that makes the dark top stable.  Depending on the details of the dark top model, $X$ might also couple other standard model fermions to their dark partners.   We will focus on the case were there is only a dark top, in which case the color triplets have 100\% branching ratio to some kind of dark top final state.  Also, though \Fig{fig:gammaplot} suggests that the preferred mass range in the simple group model would push $X$ beyond the kinematic reach of the LHC, non-thermal dark top production would allow for lighter $X$ fields.

\begin{figure}
\begin{center}
\includegraphics[scale=0.5]{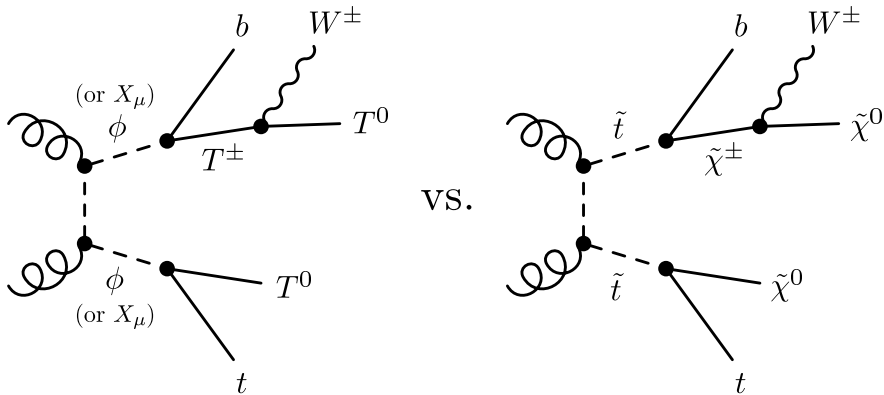}
\end{center}
\caption{Example cascade decays in the dark top model compared to supersymmetric models.  Most dark top event topologies have supersymmetric analogs, though of course the quantum numbers of the two scenarios can be different.  In the MSSM, the scalar color triplets $\tilde{t}$ carry flavor and they cascade decay to a flavorless fermion dark matter candidate $\tilde{\chi}^0$.  In the simple group dark top, the scalar color triplets $\phi$ do not carry flavor, but they cascade decay to a fermion dark matter candidate $T^0$ with third generation flavor.  One potentially large difference between the dark top and SUSY is the possible existence of $\mathbf{Z}_2$-odd spin-1 bosons $X_\mu$ as predicted in \Sec{subsec:uvfield}.  While these spin-1 fields cascade decay in much the same way as squarks, they will have different production cross sections compared to squarks for the same value of the pole mass.
\label{fig:topologies}}
\end{figure}

The $X$ fields will have strong pair production cross sections $pp \rightarrow X\overline{X}$.  Depending on their $SU(2)_L \times U(1)_Y$ quantum numbers, they will decay either as $X\rightarrow t T_i$ or $X\rightarrow b T_j$ or both.  If the mass difference between $X$ and $T_i$ is smaller than $m_{\rm top}$, decays through a charm may be enhanced.   If the dark tops are not degenerate, then an additional decay stage $T_i \rightarrow T_j + W/Z$ could be visible, where the $W/Z$ may or may not be on shell depending on the $T_i/T_j$ mass splitting.  As shown in \Fig{fig:topologies}, the topologies possible in $X$ production and decay are similar to supersymmetric models, and the discovery strategies and challenges for $X$ are similar to the discovery strategies and challenges for ordinary colored top partners.  We refer the reader to Refs.~\cite{tprime-model,tprime-study-1,Carena:2006jx,tprime-study-2,tprime-study-3,tprime-study-4,tprime-study-5,Han:2008gy} for examples of such studies.

In fact, the spin-0 $\phi$ field in \Sec{sec:simplegroup} will be nearly impossible to distinguish from a supersymmetric right-handed stop $\tilde{t}_R$ in both production and decay since both fields have the same gauge quantum numbers.\footnote{Because $\phi$ and $\tilde{t}_R$ have different flavor quantum numbers, $\phi$ could have additional decay modes to first- or second-generation quarks/dark quarks, but this assumes that these dark quarks exist and are kinematically accessible.}  In particular, the coupling
\be
\mathcal{L}_{\rm int} = \lambda_{\rm top} \phi q T_d^c, \qquad T_d^c = (T_A^c, T_B^c),
\ee
is the same as the corresponding squark-quark-higgsino $\tilde{t}_R q \tilde{H}_u$ vertex in the MSSM.   The roles of $\phi$/$\tilde{t}_R$  and $T_d^c/\tilde{H}_u$ in electroweak symmetry breaking are completely different, of course, so one would like to have some handle to distinguish these two models.

Though by no means a smoking gun, one suggestive feature of the simple group example is that the mass difference between the lightest two states is controlled by the mass of the top:
\be
m_2^2 - m_1^2 = m_{\rm top}^2 + \cdots.
\ee
In SUSY with a higgsino LSP, the lightest chargino-neutralino mass splitting is determined by higgsino-gaugino mixing and has nothing to do with the top mass.  Unfortunately, because $T_1$ is neutral but $T_2$ is charged, there is no dilepton endpoint to measure this mass difference.  On the other hand, if we pair produce $X$ and both decay as
\be
X \rightarrow b (T_2 \rightarrow W_{\rm hadronic} T_1),
\ee
then we can in principle use the methods of Refs.~\cite{Cheng:2007xv,Cho:2007qv,Barr:2007hy,Cho:2007dh} to measure $m_2^2 - m_1^2$.  If such a measurement were possible, then the fact that the mass-squared difference between dark top states is almost exactly $m_{\rm top}^2$ would be a strong indication of the dark top's role in regulating the Higgs potential.

Finally, in addition to those fields that are particular to the dark top scenario, one also may see fields associated with generic composite Higgs scenarios, such as KK gluons \cite{Pomarol:1999ad,Davoudiasl:2000wi}, $W'/Z'$ gauge bosons \cite{Burdman:2002ns,Han:2003wu}, or exotic composites \cite{Katz:2003sn,Gregoire:2008mr}.  For color neutral fields in particular, the LHC signatures of these states depend strongly on the implementation of flavor, since the production cross sections will be determined by the couplings to light quarks.  Since flavor physics could have other dramatic effects---such as changing the branching ratios of the $X$ fields or even altering the entire structure of the dark matter sector if there are additional dark partners---it would be worthwhile adding flavor to the explicit constructions of \Sec{sec:explicitads} to see how the LHC signatures change.

\section{Conclusions}
\label{sec:outlook}

There are good reasons to expect top partners and dark matter candidates to appear at the weak scale, and the dark top emphasizes the important role symmetries play in determining the properties of new BSM states.  While most BSM scenarios have colored top partners to regulate the top loop, we have seen that a combination of electroweak doublets and singlets can play the same role as long as there is an enlarged global symmetry structure.  This allows a dramatic reordering of the BSM spectrum with the dark tops being the lightest new states.

The symmetry structures of dark top models are constrained by requiring valid UV completions.  Unlike in little Higgs theories where collective symmetry breaking implied the complete absence of quadratically sensitive terms in the Coleman-Weinberg potential, in dark top models we can only make the weaker statement that the global symmetries allow for (but do not require) cancellations among various quadratically sensitive pieces.  The existence of explicit AdS$_5$ constructions is therefore an important check of the theoretical viability of the dark top proposal.

As a dark matter candidate, the dark top faces many of the same challenges as pure singlet or pure doublet dark matter models, with the added complication that the couplings of the dark top to the Higgs are fixed by the top loop cancellation requirement.    Because of the large Higgs exchange cross section, direct detection experiments constrain the dark top to be heavier than the top quark by a factor of 2 to 4 depending on the specific model.  These masses are consistent with thermal relic abundance calculations, though in the product group example, one must include the contribution from UV-sensitive four-fermion operators.  Amusingly, the dark top carries baryon number, so like the model in \Ref{Agashe:2004ci}, one could speculate about a relationship between the dark matter relic abundance and the baryon asymmetry of the universe, though the preferred dark top masses are too large to elegantly explain $\Omega_b/\Omega_{\rm DM}$.

If new colored particles are kinematically accessible at the LHC, then dark top models would have similar production and decay topologies to other BSM proposals, despite the fact that ordinary colored top partners initiate cascade decays while the dark top terminates the cascade.  In the most minimal scenario with just pair production of color triplet $X$ fields each decaying to top and dark top, it will be quite challenging to discriminate a dark top model from a supersymmetric model.  On the other hand, depending on exactly how flavor is implemented in realistic dark top models, there could be important observable differences between having flavor physics at the bottom versus top of the decay chain.

\section*{Acknowledgments}

We would like to thank Roni Harnik for detailed discussions of the symmetry structure of the dark top.  We also benefitted from conversations with Jay Wacker and Lian-Tao Wang.  J.T. is supported by a fellowship from the Miller Institute for Basic
Research in Science.  D.P. is supported under a National Science Foundation Graduate Research Fellowship.  J.T.
thanks the Aspen Center for Physics for hospitality while this work was being completed.

\appendix

\section{Dark Top Symmetry Structures}
\label{app:symmetries}

We saw explicitly that the models presented in \Sec{sec:models} regulate the Higgs potential, but it is instructive to understand to how this occurs from a symmetry point of view.

The product group example of \Sec{sec:productgroup} is the easiest to understand.  In that model, $Q^c$ preserves the $SU(3)_L$ symmetry that makes the Higgs a Goldstone boson, but $Q$ breaks $SU(3)_L$.  However, if we were to decompose $Q$ as
\be
Q = \left(\begin{array}{cccccc}q_{1r} & q_{1g} & q_{1b} & 0 & 0 & 0 \\q_{2r} & q_{2g} & q_{2b}  & 0 & 0 & 0 \\T'_A & T'_B & T'_C & 0 & 0 & 0\end{array}\right),
\ee
then $Q$ would respect the $SU(3)_L$ symmetry that protected the Higgs mass.  Because $T'_i$ and $T_i$ are in an $SU(6)_C$ multiplet, the radiative corrections from $T_i$ are identical to the radiative correction from $T_i'$, so the original decomposition of \Eq{eq:decompose63} has an effective ``twisted'' $SU(3)_L$ symmetry, at least as far as the quadratically divergent one-loop corrections are concerned.  This kind of twisting is familiar from twin Higgs \cite{Chacko:2005pe,Chacko:2005un} and folded supersymmetric \cite{Burdman:2006tz} models.

The simple group example of \Sec{sec:simplegroup} works in a similar way, but the details are more confusing because $Q$ is a symmetric tensor.  Imagine there were two $SU(6)$ symmetries $SU(6)_{L,R}$, two sets of $Q$, two sets of $Q^c$, and two sets of $\Phi$ with transformation properties
\be
Q_L: (\mathbf{\bar{6}}, \mathbf{\bar{6}}), \quad Q^c_L : (\mathbf{1}, \mathbf{6}), \quad \Phi_L: (\mathbf{6},\mathbf{1}),
\ee
\be
Q_R: (\mathbf{\bar{6}}, \mathbf{\bar{6}}), \quad Q^c_R : (\mathbf{6}, \mathbf{1}), \quad \Phi_R: (\mathbf{1},\mathbf{6}).
\ee
Assuming a $\mathbf{Z}_2$ symmetry between $SU(6)_L$ and $SU(6)_R$, we can write down the Yukawa couplings
\be
\label{eq;yukawa65doubled}
\mathcal{L}_{\rm Yukawa} = \sqrt{2} \lambda_{\rm top} \left[Q_L \Phi_L Q^c_L + Q_R \Phi_R Q^c_R  \right].
\ee
If we decompose $Q_i$ as (the subscript on the matrix applies to every field)
\be
Q_L = \left(\begin{array}{ccccccc} 0 & 0 & 0 & 0 & 0 & 0  \\0 & 0 & 0 & 0 & 0 & 0  \\0 & 0 &  0& 0&0 & 0 \\q_{1r} & q_{1g} & q_{1b} & 0 & 0 &  0 \\q_{2r} & q_{2g} & q_{2b} & 0 & 0& 0 \\T_A' & T_B' & T_C' & 0 & 0 & 0 \end{array}\right)_L, \qquad Q_R = \left(\begin{array}{ccccccc} 0 & 0 & 0 & q_{1r} & q_{2r} & T_A'  \\0 & 0 & 0 & q_{1g} & q_{2g} & T_B'  \\0 & 0 &  0& q_{1b} & q_{1b} & T_C' \\0&0&0&0&0&0 \\0&0&0&0&0&0\\0&0&0&0&0&0 \end{array}\right)_R,
\ee
then there is an enhanced $SU(3)_L$ symmetry that protects the Higgs mass in $\Phi_L$ and an enhanced $SU(3)_R$ symmetry that protects the Higgs mass in $\Phi_R$.  We can ``twist'' that $SU(3)^2$ symmetry by decomposing $Q_i$ as
\be
Q_L = \left(\begin{array}{ccccccc} 0 & 0 & 0 & 0 & 0 & 0  \\0 & 0 & 0 & 0 & 0 & 0  \\0 & 0 &  0& 0&0 & 0 \\q_{1r} & q_{1g} & q_{1b} & 0 & 0 &  0 \\q_{2r} & q_{2g} & q_{2b} & 0 & 0& 0 \\0&0&0&T_A & T_B & T_C \end{array}\right)_L, \qquad Q_R = \left(\begin{array}{ccccccc} 0 & 0 & 0 & q_{1r} & q_{2r} & 0  \\0 & 0 & 0 & q_{1g} & q_{2g} & 0 \\0 & 0 &  0& q_{1b} & q_{1b} & 0 \\0&0&0&0&0&T_A \\0&0&0&0&0&T_B\\0&0&0&0&0&T_C \end{array}\right)_R.
\ee
If we now break the $SU(6)_L \times SU(6)_R$ symmetry down to the diagonal, we can identify $\Phi_L$ with $\Phi_R$ and give infinite vector-like masses to the components of
\be
Q_{L}^{ij} - Q_{R}^{ji}, \qquad Q^c_{Li} - Q^c_{Ri}.
\ee
Integrating out these anti-symmetric fermion components, the Yukawa coupling in \Eq{eq;yukawa65doubled} does indeed reduce to \Eq{eq:Yukawa65} with the decomposition of \Eq{eq:decompose65}, once we go to canonical normalization for the fields.

\section{Deconstructed Radiative Potential}
\label{sec:deconstructed}

\begin{figure}
\begin{center}
\includegraphics[scale=0.35]{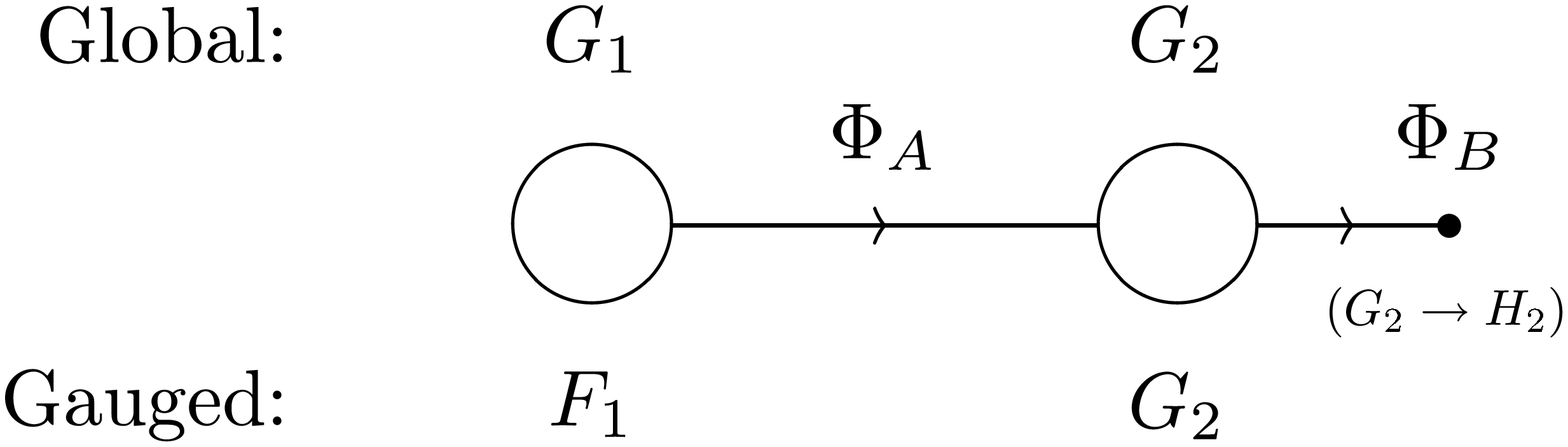}
\end{center}
\caption{A two-site deconstruction of the AdS$_5$ model from \Fig{fig:adsexample}.  There are two copies of the $G$ global symmetry, and $G_1 \times G_2$ is broken to the diagonal subgroup through the linear sigma field $\Phi_A$.   On the first site, an $F_1$ subgroup of $G_1$ is gauged.  On the second site, all of $G_2$ is gauged, but $G_2$ is spontaneously broken to $H_2$ through the linear sigma field $\Phi_B$.  The spin-0 and spin-1 spectrum of this theory matches the lowest KK modes in the AdS$_5$ construction.
\label{fig:deconstructedexample}}
\end{figure}

In \Sec{subsec:adshiggs}, we calculated the one-loop radiative Higgs potential from fermion loops in an explicit AdS$_5$ UV completion of the dark top.  We found that for the $SU(6)/SU(5)$ simple group example from \Sec{sec:simplegroup} there was an effective cancellation of the top quadratic divergence, whereas for the inviable $SU(7)/SU(6)$ example from \Sec{sec:tension} the quadratic divergence was effectively unregulated.  To understand better why this is happening, it is instructive to redo the calculation in a two-site deconstructed \cite{ArkaniHamed:2001ca} version of the AdS$_5$ theory.

For simplicity, we will describe only the $SU(6)/SU(5)$ model since the generalization to $SU(7)/SU(6)$ is straightforward.  We will use a linear sigma model language and turn off hypercharge.  As shown in \Fig{fig:deconstructedexample}, the deconstructed theory has two copies of the global symmetry group $SU(6)_1$ and $SU(6)_2$.  The linear sigma field $\Phi_A$ transforms as a $(\mathbf{6},\mathbf{\overline{6}})$ of $SU(6)_1 \times SU(6)_2$ and breaks the symmetry to the diagonal $SU(6)_V$.  The linear sigma field $\Phi_B$ transforms as a $\mathbf{6}$ of $SU(6)_2$ and breaks $SU(6)_2 \rightarrow SU(5)_2$.  The standard model $SU(3)_C \times SU(2)_L $ subgroup of $SU(6)_1$ is gauged and all of $SU(6)_2$ is gauged.  The standard model gauge bosons are contained in the relevant generators of the diagonal $SU(6)_V$.

In unitary gauge, the sigma fields take the form
\be
\Phi_A = e^{i\Pi/ f} \langle \Phi_A \rangle , \qquad \Phi_B = e^{i \Pi/f} \langle \Phi_B \rangle,
\ee
where $\Pi$ is the uneaten Goldstone boson matrix, and the effective pion decay constant is $f = \sqrt{f_A^2 + f_B^2}$.  The vevs are
\be
\langle \Phi_A \rangle = \diag(f_A,f_A,f_A,f_A,f_A,f_A), \qquad \langle \Phi_B \rangle = \left(\begin{array}{c}0 \\0 \\ 0 \\0 \\0 \\f_B \end{array}\right).
\ee

The deconstructed theory has four fermion fields $Q_1$, $Q_1^c$, $Q_2$, $Q_2^c$.  Under $SU(6)_1 \times SU(6)_2$, they transform as
\be
Q_1 \colon (\mathbf{\overline{6}},\mathbf{1}), \qquad
Q_1^c \colon (\mathbf{21},\mathbf{1}),  \qquad
Q_2 \colon (\mathbf{1},\mathbf{\overline{21}}),  \qquad
Q_2^c \colon (\mathbf{1},\mathbf{6}).
\ee
Note that $Q_2$ and $Q_2^c$ are in the same representations as the original $Q$ and $Q^c$ in the low energy theory.  These representations allow the following Yukawa couplings
\be
\mathcal{L}_{\rm Yukawa} = \frac{m_c}{f_A} Q_1 \Phi_A Q_2^c + \frac{m}{f_A^2} Q_1^c \Phi_A^\dagger \Phi_A^\dagger Q_2 + \sqrt{2} \lambda_{\rm top} Q_2 \Phi_B Q_2^c,
\ee
where $\Phi_A$ appears twice in the second term because $Q_2$ and $Q_1^c$ are symmetric tensors.  We take $Q_2$ and $Q_2^c$ to be complete multiplets of $SU(6)_2$, while $Q_1$ and  $Q_1^c$ are incomplete multiplets of $SU(6)_1$ with the \emph{opposite} components zeroed out compared to \Eq{eq:decompose65}.  This way we recover the same fermion content as the low energy models in the $m, m_c \rightarrow \infty$ limit.  In the $SU(6)/SU(5)$ model, all of the components of $Q_1$ are zeroed out, but we keep $Q_1$ in the analysis since the $SU(7)/SU(6)$ analog has one non-zero component.  Because we are using only a two-site model, there is no fermion mixing, so $\lambda_{\rm top}$ is the physical top Yukawa.

It is now straightforward to use the ordinary Coleman-Weinberg potential from \Eq{eq:colemanweinberg} to calculate the one-loop contribution to the fermion masses.  Ignoring numerical coefficients and working in the limit that $f \ll m,m_c$, the radiative Higgs mass is proportional to
\be
\label{eq:65deconhiggsmassestimate}
SU(6)/SU(5): \qquad \delta m^2_h \propto \lambda_{\rm top}^2 f^2 \log \frac{f}{m}.
\ee
We see that the Higgs mass depends quadratically on $f$ but only logarithmically on the ``cutoff'' $m$, showing that the low energy quadratic divergence is indeed cancelled.  Note that only $m$ and not $m^c$ appears in the theory, since $Q_1$ does not contribute at all to the potential.

We can do the same calculation in the $SU(7)/SU(6)$ model, and the leading term is
\be
SU(7)/SU(6): \qquad \delta m^2_h \propto \lambda_{\rm top}^2 \left[m^2 \log \frac{m}{\Lambda} - m_c^2 \log \frac{m_c}{\Lambda}  \right],
\ee
where $\Lambda$ is the explicit cutoff from the Coleman-Weinberg potential, and corresponds roughly to the mass of the second KK mode in AdS space.  There is also a subleading term that has the same form as \Eq{eq:65deconhiggsmassestimate}.  If $m = m^c$, then the leading term cancels, leaving only the subleading contribution.  However, there is no reason to expect $m$ and $m^c$ to be the same since there is no symmetry relating them.  On the other hand, the masses of the first KK modes in AdS space are not that different for different fermion species, so it makes sense to expand in $m^c - m = \delta m$.  The leading term from this expansion is
\be
SU(7)/SU(6) \mbox{ (small $\delta m$)}: \qquad \delta m^2_h \propto \lambda_{\rm top}^2 m \, \delta m \log \frac{m}{\Lambda},
\ee
so instead of quadratic growth in $m$ there is only linear growth.  Still, in typical AdS constructions $m \, \delta m$ is larger than $f^2$, so the top quadratic divergence is better regulated in the $SU(6)/SU(5)$ model compared to the $SU(7)/SU(6)$ model.

\end{document}